\documentclass[10pt]{article}

\usepackage{amsmath}
\usepackage{amssymb}

\usepackage{graphicx}
\usepackage{ragged2e}

\usepackage{cite}

\usepackage{color} 
\usepackage{multirow}
\usepackage{array}
\usepackage{natbib}

\newcommand{\one}{($i$) }
\newcommand{\two}{($ii$) }
\newcommand{\three}{($iii$) }
\newcommand{\four}{($iv$) }
\newcommand{\five}{($v$) }
\newcommand{\six}{($vi$) }

\newcommand{\onel}{($A$)}
\newcommand{\twol}{($B$)}
\newcommand{\threel}{($C$)}
\newcommand{\fourl}{($D$)}
\newcommand{\fivel}{(\textbf{E})}
\newcommand{\sixl}{(\textbf{F})}

\newcommand{\sm}{\emph{Appendix}}

\usepackage[labelfont=bf,labelsep=period,justification=raggedright]{caption}

\makeatletter
\renewcommand{\@biblabel}[1]{\quad#1.}
\makeatother

\date{}

\begin{document}

\title{The influence of transcription factor competition on the relationship between occupancy and affinity} 
\author{Nicolae Radu Zabet$^{1,2,\ast}$, Robert Foy$^{1,2}$ and Boris Adryan$^{1,2,\dagger}$    \\
\mbox{}\\
\footnotesize $^1$Cambridge Systems Biology Centre, University of Cambridge, Tennis Court Road, Cambridge CB2 1QR, UK\\
\footnotesize $^2$Department of Genetics, University of Cambridge, Downing Street, Cambridge CB2 3EH, UK\\
\footnotesize $^\ast$Email: n.r.zabet@gen.cam.ac.uk\ \ \footnotesize $^\dagger$Email: ba255@cam.ac.uk}

\maketitle

\begin{abstract}
Transcription factors (TFs) are proteins that bind to specific sites on the DNA and regulate gene activity. Identifying where TF molecules bind and how much time they spend on their target sites is key for understanding transcriptional regulation. It is usually assumed that the free energy of binding of a TF to the DNA (the affinity of the site) is highly correlated to the amount of time the TF remains bound (the occupancy of the site). However, knowing the binding energy is not sufficient to infer actual binding site occupancy. This mismatch between the occupancy predicted by the affinity and the observed occupancy may be caused by various factors, such as TF abundance, competition between TFs or the arrangement of the sites on the DNA. We investigated the relationship between the affinity of a TF for a set of binding sites and their occupancy. In particular, we considered the case of lac repressor (lacI) in \emph{E.coli} and performed stochastic simulations of the TF dynamics on the DNA for various combinations of lacI abundance in competition with TFs that contribute to macromolecular crowding. Our results showed that for medium and high affinity sites, TF competition does not play a significant role in genomic occupancy, except in cases when the abundance of lacI is significantly increased or when a low-information content PWM was used. Nevertheless, for medium and low affinity sites, an increase in TF abundance (for both lacI or other molecules) leads to an increase in occupancy at several sites.

\end{abstract}

\section{Introduction}

A powerful key to understanding transcriptional regulation is the amount of time a regulatory binding site is occupied by a cognate transcription factor (TF). In particular, this `occupancy' measure can be used to infer relative amounts of transcription of the target gene, and is therefore a more powerful comparative tool than simple sequence searches for `preferred binding sites'. Transcription factors have specific affinities for each site on the DNA (computed from the binding energy between the TF protein and the DNA molecule at the target site) and it is often na{\"i}vely assumed that this affinity is sufficient to predict the actual occupancy of TFs bound to the DNA  \citep{segal_2009}. However, recent studies have demonstrated that affinity alone is not always sufficient to accurately predict TF occupancy \citep{kaplan_2011}.

Previous studies have shown that TF abundance can account for the correlation between the normalised affinity and normalised occupancy (``normalised" here refers to setting the maximum observed values to $1$) \citep{hippel_1986,berg_1987,gerland_2002,djordjevic_2003,roider_2007,zhao_2009}, in the sense that  increasing TF abundance increases the number of occupied sites and that those additional sites are of decreasing affinity. This result was explained by the fact that, once the high affinity sites get close to saturation, TF molecules will spend more time bound to lower affinity sites. However, in those studies the spatial organisation of sites on the DNA was disregarded. Such an assumption should predict occupancy for \emph{in vitro} experiments such as SELEX or PBM \citep{stormo_2010}, (where there are only short DNA sequences and one TF species), whilst  in  \emph{in vivo} studies, could lead to biased predictions.

A popular approach to estimate occupancy is the statistical thermodynamics framework. This method computes the probability that, at equilibrium, one encounters a specific configuration of TF molecules on the DNA \citep{ackers_1982,bintu_2005_model,bintu_2005b,sadka_2009}. A number of studies consider a uniform affinity landscape for TFs or other DNA-binding proteins and focus on the occupancy of a single site (or a few sites) in the context of a genome with otherwise constant affinity \citep{ackers_1982,bintu_2005_model,bintu_2005b,sadka_2009}. However, TFs display a distribution of affinities to the DNA \citep{stormo_2000,gerland_2002} and, thus, the assumption of a uniform landscape becomes restrictive (and can lead to biases in the results). \citet{wasson_2009} considered non-uniform affinity landscapes and investigated the relationship between the abundance of DNA-binding proteins and their occupancy using a statistical thermodynamics model. Their results confirmed that, when increasing TF abundance, low affinity sites display higher occupancy than that which would be predicted by affinity alone. Furthermore, the addition of other DNA-binding proteins (histones in their case) leads to an overall reduction in occupancy of the TFs of interest. Similarly, \citet{kaplan_2011} applied a combination of a hidden Markov model and a thermodynamic framework and discovered that TF competition does not influence the observed occupancy significantly (at least in the case of their system). Nevertheless, they considered only the competition between various TF species and did not alter the abundance of their TFs of interest (they used the actual TF abundance that was experimentally measured). 

The main assumption of the statistical thermodynamic framework is that the system reaches equilibrium and the transient time (the time to reach equilibrium) is negligible \citep{segal_2009}. Nevertheless, there is still no proof that, in the case of the TF search process, equilibrium exists or is reached fast enough to not affect the average behaviour. We use a stochastic simulation of the process by which a TF `searches' for it's regulatory binding site by first binding non-specifically to the DNA and then performing a one-dimensional random walk before eventually unbinding. This combination of binding/unbinding to/from the DNA and one-dimensional random walk is known as a \emph{facilitated diffusion mechanism} \citep{berg_1981} and it is evident that such a process is taking place inside the cell \citep{elf_2007,hammar_2012}. The physical advantage of facilitated diffusion over a purely three-dimensional diffusion or a purely one-dimensional random walk is a more rapid target site location; see \citep{zabet_2012_review}. Simulating facilitated diffusion can overcome some of the limitations of the statistical thermodynamics model by allowing `exact' \emph{in silico} measurement of the average occupancy of TF binding sites under various parametrisations of the cellular state (e.g. concentrations of DNA binding proteins), some of which will give rise to deviations from the predictions offered by the statistical thermodynamics model. For example, \citet{chu_2009} demonstrate such deviations when they model TFs as having non-uniform affinity landscapes.

Here, we used a stochastic simulator that models the facilitated diffusion mechanism and studied the properties of a complete continuous DNA sequence (from the genome of \emph{E.coli} K-12 \citep{riley_2006}) being bound by both a cognate TF species (lacI in our case) and a non-cognate TF species (aimed to model the presence of other proteins on the DNA which contribute to crowding on the DNA) \citep{zabet_2012_grip,zabet_2012_model}. This scenario mimics the behaviour of TF molecules in a live cell performing facilitated diffusion in the search for their target sites. The TF molecules will not only compete with other molecules bound to the DNA for sites, but during the one-dimensional random walk on the DNA, they will slide or hop to nearby sites \citep{mirny_2009} and also bypass other bound molecules \citep{kampmann_2004,hedglin_2010} which act as obstacles and create boundary effects \citep{segal_2009}. 

Our results confirm that the addition of non-cognate TFs reduces the \emph{absolute} occupancy of cognate TF binding sites, while their \emph{relative} occupancy is influenced at relatively few (in the order of tens) low and medium affinity sites, and is unaffected at high affinity sites. That is, for low affinity (``non-specific") and medium affinity sites, the addition of non-cognate TFs leads to significant differences between the predicted  relative occupancy based on affinity (which we call affinity derived occupancy, or ADO) and the relative occupancy measured by stochastic simulation (which we call simulation derived occupancy, or SDO) at several sites, whilst for high affinity sites this relative binding pattern is unaffected. While the mismatch associated with low affinity sites should have little or no influence on gene regulation (unless the cognate TF molecules change conformation when bound to a functional high affinity site \citep{marcovitz_2011}), this may provide an explanation for the noise structure in actual genomic profiles of TF occupancy (e.g. ChIP data). 

We further found that differences between ADO and SDO at medium and high affinity sites can arise if the cognate TF abundance is significantly increased or if the information content of the PWM is low. However, for normal bacterial TF abundances (usually in the range of $10-100$ copies \citep{wunderlich_2009}), PWM information content \citep{stormo_1998,wunderlich_2009} and DNA sizes (e.g., $4.6\ Mbp$ \citep{riley_2006}), the differences between the SDO and ADO are negligible and binding energies are good indicators of occupancy. Nevertheless, in the case of eukaryotic systems, their high TF abundances ($> 10^4$ copies \citep{biggin_2011}), their lower information content motifs \citep{wunderlich_2009} and the amount of \emph{accessible} DNA suggest that significant differences between ADO and SDO are likely to occur. Nevertheless, this increase in occupancy generated by the high abundance of cognate TFs can be reduced, to a certain degree, by a high abundance of non-cognate TF molecules in the system.

\section{Materials and Methods}
We use GRiP \citep{zabet_2012_grip} to simulate facilitated diffusion of DNA-binding proteins around the DNA, which allows parametrisation with affinity data and measures site occupancy. Briefly, GRiP performs event driven stochastic simulations \citep{gillespie_1976,gillespie_1977} of all molecules in the cell which are explicitly represented. Molecules perform both a three-dimensional diffusion in the cytoplasm (nucleoplasm in the case of eukaryotic cells) and a one-dimensional random walk on the DNA. The three-dimensional diffusion is modelled implicitly by simulating the Chemical Master Equation. This approach was shown to display negligible error if fast rebinding to the DNA is also modelled \citep{zon_2006}, and, in GRiP, fast rebinding is modelled through hopping mechanism of TFs on the DNA. In addition, the model implements steric hindrance, in the sense that any base pair cannot be covered by two TFs  simultaneously \citep{hermsen_2006}. The complete set of parameters for the model were previously presented in \citep{zabet_2012_model} and can be found in \sm \ref{seq:appendixTFparams}.

\begin{figure}[!ht]
\begin{center}
\includegraphics[scale=0.5,angle=270]{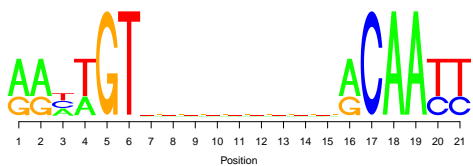}
\end{center}
\caption[lacI sequence logo]{\justifying{\bf lacI sequence  logo} The canonical lacI motif as generated from the three known high affinity sites \citep{zabet_2012_subsystem}.}
\label{fig:lacI3g}
\end{figure}

In this study, we consider the case of lac repressor (lacI) TF in \emph{E.coli}, with an association rate to the DNA of $k^{\textrm{assoc}}_{\textrm{lacI}}=2400\ s^{-1}$ \citep{zabet_2012_subsystem} and a specificity as modelled by the PWM in Figure \ref{fig:lacI3g}.

In addition to lacI, the system explicitly represents non-cognate molecules in order to model macromolecular crowding. Each non-cognate molecule covers $46\ bp$ of DNA and is allowed to perform the facilitated diffusion mechanism in a similar way to cognate molecules  \citep{zabet_2012_model}. We consider five levels of crowding, namely: \one $0\%$ ($TF_{\textrm{nc}}^{0}=0$), \two $9\%$ ($TF_{\textrm{nc}}^{0.09}=10^{4}$ and $k^{\textrm{assoc}}_{\textrm{nc}}=2000\ s^{-1}$), \three $26\%$ ($TF_{\textrm{nc}}^{0.26}=3\times 10^{4}$ and $k^{\textrm{assoc}}_{\textrm{nc}}=2571\ s^{-1}$), \four $42\%$ ($TF_{\textrm{nc}}^{0.42}=5\times 10^{4}$ and $k^{\textrm{assoc}}_{\textrm{nc}}=3600\ s^{-1}$) and \five $55\%$ ($TF_{\textrm{nc}}^{0.55}=7\times 10^{4}$ and $k^{\textrm{assoc}}_{\textrm{nc}}=6000\ s^{-1}$). Note that, with the exception of the first case (no crowding on the DNA), all cases display crowding which is within biologically plausible values ($10\%$ to $50\%$ \citep{flyvbjerg_2006}).

Before proceeding to investigate the relationship between affinity derived occupancy (ADO) and simulation derived occupancy (SDO), we first need to describe the methods used to estimate these parameters. ADO is computed using the average time a TF molecule spends bound at a certain position on the DNA as derived from an approximation of the binding energy (which is itself calculated from PWM score); see equation (3) in \citep{zabet_2012_model}. Briefly, the affinity predicted occupancy of a TF bound at the $j^{th}$ nucleotide on the DNA is given by
\begin{equation}
\tau_{\textrm{lacI}}^{j} = \tau_{\textrm{lacI}}^{0} \exp{\left[ \frac{1}{K_BT}\left(- E_{\textrm{lacI}}^{j}\right)\right]}
\end{equation}
where $\tau_{\textrm{lacI}}^{0}$ is the average waiting time when bound at $O_1$ site, $E_{\textrm{lacI}}^{j}$ is the binding energy at position $j$ (which is equal to $E_{\textrm{lacI}}^{j}=-w{\textrm{lacI}}^{j}$, where $w{\textrm{lacI}}^j$ is the lacI PWM score at the $j^{th}$ nucleotide),  $K_B$ is the Boltzmann constant and $T$ the temperature. In \citep{zabet_2012_subsystem}, we computed $\tau_{\textrm{lacI}}^{0}=1.18e^{-06}$. 

All ADO vs SDO plots consider log values that are normalised to the maximum ADO or SDO, respectively. For example, in the case of affinity predicted occupancy, we plot:
\begin{equation}
\log \left( \frac{\tau_{\textrm{lacI}}^{j}}{\displaystyle \max_i \{ \tau^i_{lacI} \}}  \right)
\end{equation}

While ADO is computed directly from the PWM (\emph{a priori} to the simulations) the SDO (simulation derived occupancy) is based on the results of our stochastic simulations. There are several ways in which the SDO can be estimated and in the following section we compare these approaches to justify our choice. 

\subsection{Measuring the occupancy}

There are three methods to estimate the observed occupancy, namely:

\begin{enumerate}
\item	\emph{Ensemble average} - Perform a set of $X$ stochastic simulations with identical parameters, each running for a time interval $T_s$ (chosen as adequate to reach a stationary behaviour) and  record the position of each molecule at the end of the simulation. Using these $X$ sets of positions, measure the occupancy by computing the average amount of time the TF spends at each position \citep{kaplan_2011}. [Note: this is effectively the result obtained from a ChIP experiment: the mean behaviour within an ensemble of cells.] 
\item	\emph{Time average} - Observe a single  system for a much longer time interval $T_l$ and compute the occupancy as the average amount of time the TF spends at each position \citep{zabet_2012_model}. The time average can take less time to compute and, consequently, is an appealing method to estimate occupancy. In live cells, the activity state of a gene is related to the proportion of time the regulatory region is occupied and, thus, the time average may be a better indicator for biological relevance than ensemble average \citep{zabet_2012_review}. Nevertheless, if one wants to replicate the result of ChIP experiments, then the ensemble average is more appropriate. 
\item \emph{Hybrid average} - Perform a set of $X$ stochastic simulations for a long time interval $T_l$. For each simulation calculate the time average occupancy and then perform an ensemble average over all time averages. At the population level, there is an ensemble average over the behaviour of all cells, thus the hybrid average is a good indicator of the occupancy when investigating gene regulation at population level. 
\end{enumerate}

The ergodic theorem assumes that the time average for long time intervals equals the ensemble average. However, the ergodicity assumption breaks down in certain cases (e.g. the time average differs from the ensemble average in multi-stable systems \citep{gillespie_2000}). Thus, we need to investigate under what conditions the ergodicity assumptions break down within our system.

\begin{figure}[!ht]
\begin{center}
\includegraphics[scale=0.55,angle=270]{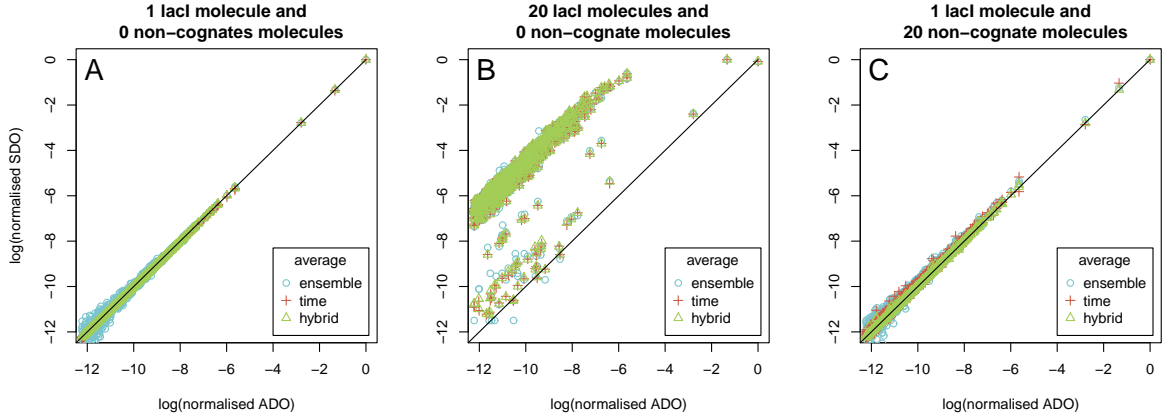}
\end{center}
\caption[Comparison between the ensemble, time and hybrid averages of SDO in a crowded environments]{\justifying{\bf Comparison between the ensemble, time and hybrid averages of SDO in a crowded environment.} We considered  $1\ Kbp$ of DNA, which contains the $O_1$ site (the strongest known binding site for lacI, which is located at  position $365,547-365,567$ on the \emph{E.coli} K-12 genome) and: \onel\ $1$ lac repressor molecule and $0$ non-cognate molecules, \twol\ $20$ lac repressor molecules and $0$ non-cognate molecules and \threel\ $1$ lac repressor molecule and $20$ non-cognate molecules.  We plotted the sites that have a binding energy at least $30\%$ of the highest value ($577$ strongest sites). \onel\ The ensemble average is computed from $X=2\times 10^{6}$ independent simulations [blue circles]; the time average  is computed by running the simulations for $T_l=3000\ s$ [red crosses]; and the hybrid average  is computed by running $X=40$ independent simulations for $T_l=3000\ s$ [green triangles]. \twol\ The ensemble average is computed from $X=1\times 10^{5}$ independent simulations (blue circles); the time average is computed by running the simulations for $T_l=150\ s$ [red crosses]; and the hybrid average is computed by running $X=40$ independent simulations for $T_l=150\ s$ [green triangles].  \threel\ The ensemble average is computed from $X=2\times 10^{6}$ independent simulations [blue circles]; the time average is computed by running the simulations for $T_l=3000\ s$ [red crosses]; and the hybrid averageis computed by running $X=40$ independent simulations for $T_l=3000\ s$ [green triangles]. Table \ref{tab:methodsTtests} shows that the three measures for SDO appear to have the same mean.}
\label{fig:ergodicityHybridScatterplot}
\end{figure}

Figure \ref{fig:ergodicityHybridScatterplot}\onel\ confirms that the time average, hybrid average and ensemble average measures for SDO produce similar results. In this case, the system consists of a DNA molecule and one lacI TF and zero non-cognates. In addition, one can observe that all measures for SDO display negligible differences from ADO.

By increasing the copy number of the TF, the ensemble average and time average diverge. Figure~\ref{fig:ergodicityHybridScatterplot}\twol\ models 20 lacI molecules and zero non-cognates, and it is clear that in some cases the time average values (red crosses) diverge from their associated ensemble average values (blue circles) and hybrid average values (green triangles). The more dramatic effect, however, is the significant deviation of SDO from ADO for all three measures. This shows that for significantly increased TF copy number, whilst the ergodicity assumption has begun to break down, the differences introduced are insignificant compared to the increased SDO observed at a large number of sites.

The case of increased crowding on the DNA, as modelled by the addition of non-cognate TFs, is shown in Figure~\ref{fig:ergodicityHybridScatterplot}\threel. Here the cognate abundance is kept fixed to one molecule, while 20 non-cognates are modelled. The figure shows that a significant increase in the number of non-cognates has a negligble effect on all three measures of SDO. 

Table \ref{tab:methodsTtests} shows that in the case of naked DNA and one molecule of lacI, the three measurements for SDO (ensemble, time and hybrid averages) have approximately the same mean. However, molecular crowding on the DNA leads to deviations between ensemble and hybrid averages. In particular, in the case of high abundance of cognate TFs - 20 molecules of lacI - we observed a mean increase of $\sim 33\%$ in the hybrid average compared to the ensemble average, while in the case of high abundance of non-cognate TFs - 20 non-cognate molecules - we observed a decrease of $\sim 14\%$ in the hybrid average compared to the ensemble average. In addition, in \sm \ref{seq:appendixErgodicity} we show that, when the simulation time is increased, the mean ratio of hybrid and ensemble averages tends to $1$ and the deviations from the mean are reduced. 

\begin{table}[!ht]
\centering
  \begin{tabular}{|c|>{\centering\arraybackslash}p{1.5cm}|>{\centering\arraybackslash}p{1.5cm}|>{\centering\arraybackslash}p{1.5cm}|>{\centering\arraybackslash}p{1.5cm}|>{\centering\arraybackslash}p{1.5cm}|>{\centering\arraybackslash}p{1.5cm}|>{\centering\arraybackslash}p{1.5cm}|>{\centering\arraybackslash}p{1.5cm}|}
    \hline
    & \multicolumn{2}{|c|}{\textbf{1 lacI}} &  \multicolumn{2}{|c|}{\textbf{20 lacI}} &  \multicolumn{2}{|c|}{\textbf{1 lacI}}\\
    & \multicolumn{2}{|c|}{\textbf{0 non-cognates}} &  \multicolumn{2}{|c|}{\textbf{0 non-cognates}} &  \multicolumn{2}{|c|}{\textbf{20 non-cognates}}\\ \hline
      & $mean$ &  $p.value$ &  $mean$ &  $p.value$ & $mean$ &  $p.value$  \\ \hline
    $log(time/ensemble)$ & $-0.0132$ & $0.1687$ & $-0.0148$ & $0.1546$ & $0.0788$ & $2.65e^{-13}$\\ \hline
	$log(hybrid/ensemble)$ & $0.0221$ & 0.0212 & $0.0112$ & 0.2800 & -0.1513 & $2.21e^{-51}$\\ \hline
  \end{tabular}
\caption[mean and t-test p-values of $log(time/ensemble)$ and $log(hybrid/ensemble)$ averages of SDO for three levels of crowding.]{\justifying\emph{mean and t-test p-values of $log(time/ensemble)$ and $log(hybrid/ensemble)$ averages of SDO for three levels of crowding.} The table shows the effect of crowding for different measures of occupancy. The three measures are \emph{time average}, \emph{ensemble average} and \emph{hybrid average}. The system model is as in Figure~\ref{fig:OccupancyAffinityAbundanceInformationSignificant}. The log ratios of (\emph{time/ensemble}) and (\emph{hybrid/ensemble}) show significant deviations from zero as measured by a standard one-sample t-test in the case of 1 lacI and 20 non-cognates. This demonstrates that the ergodic theorem does not hold for this level of crowding as measured by the model.}
\label{tab:methodsTtests}
\end{table}

Due to the fact that we are interested in genomic occupancy of TFs that are involved in the regulation of transcription and that, in particular, we are interested in cell population results, we use the hybrid average in all subsequent calculations within this manuscript. Nevertheless, it should be noted that using any of the three methods will lead to similar results.

\subsection{System size reduction}
Our results are obtained by simulating TF occupancy on the $100\ Kbp$ of the \emph{E.coli} K-12 genome \citep{riley_2006} (the DNA locus [300000, 400000]), roughly centered around the $O_1$ site (the most strongly bound site for lacI). In \citep{zabet_2012_subsystem}, we proposed two models that are required to adapt the parameters of the subsystem, namely: \one copy number model and \two association rate model. The former is easier to implement, but can be applied only to highly abundant TFs, while the latter requires an extra set of simulations, but can be applied to TFs with any abundance. Due to the fact that non-cognate TFs are highly abundant in our system, we applied the copy number model to simulate the non-cognate TFs. This leads to the association rate between non-cognate TFs and DNA being unaffected, but the abundances of non-cognate TFs changing to: \one $TF_{\textrm{nc}}^{0}=0$ for $0\%$ crowding, \two $TF_{\textrm{nc}}^{0.09}=216$ for $9\%$ crowding, \three $TF_{\textrm{nc}}^{0.26}=647$ for $26\%$ crowding, \four$TF_{\textrm{nc}}^{0.42}=1078$ for $42\%$ crowding and \five $TF_{\textrm{nc}}^{0.55}=1509$ for $55\%$ crowding. Note that, in this manuscript, crowding refers to the percentage of the simulated DNA covered by DNA-binding proteins.

For lacI, we considered four abundances, namely:  $1$, $10$, $100$, $1000$. Due to the lower copy number, we used the association rate approach to adjust the parameters of the full system to the subsystem. This leads to the copy number of lacI being unaffected, but its association rate changing from $k^{\textrm{assoc}}_{\textrm{lacI}}=2400\ s^{-1}$ \citep{zabet_2012_subsystem} to the values listed in Table \ref{tab:paramsSubsystem}. In \sm \ref{seq:appendixSubsystem}, we plotted the proportion of time spent on the DNA (which is required when computing the association rate) and also confirmed that our system size reduction method leads to a system behaviour that deviates only negligibly from the behaviour of the full system.

\begin{table}[!ht]
 \centering
    \begin{tabular}{|>{\centering\arraybackslash}m{20mm}|>{\centering\arraybackslash}m{20mm}|>{\centering\arraybackslash}m{20mm}|>{\centering\arraybackslash}m{20mm}|>{\centering\arraybackslash}m{20mm}|} \hline
	 \textbf{covered DNA} & $\overline{k}^{\textrm{assoc}}_{1\textrm{lacI}}\ s^{-1}$ & $\overline{k}^{\textrm{assoc}}_{1\textrm{lacI}}\ s^{-1}$ & $\overline{k}^{\textrm{assoc}}_{1\textrm{lacI}}\ s^{-1}$ & $\overline{k}^{\textrm{assoc}}_{1\textrm{lacI}}\ s^{-1}$ \\  \hline
       $0\%$ & $4.19$ & $4.04$ & $4.11$ & $4.19$ \\ \hline
    	$9\%$ & $4.58$ & $4.63$ & $4.67$ & $4.74$\\ \hline
    	$26\%$& $6.11$ & $6.10$ & $6.19$ & $6.32$\\ \hline
   	 $42\%$& $8.63$ & $8.76$ & $8.73$ & $8.88$\\ \hline
   	 $55\%$  & $13.15$ & $13.05$ & $13.06$ & $13.26$\\ \hline
    \end{tabular}
\caption[The association rate of lacI in the $100\ Kbp$ subsystem for various crowding levels on the DNA]{\justifying\emph{The association rate of lacI in the $100\ Kbp$ subsystem for various crowding levels on the DNA}The over bar is used to denote the corresponding parameters in the subsystem.}
\label{tab:paramsSubsystem}
\end{table}

\section{Results}
\begin{figure}[!ht]
\begin{center}
\includegraphics[scale=0.7,angle=270]{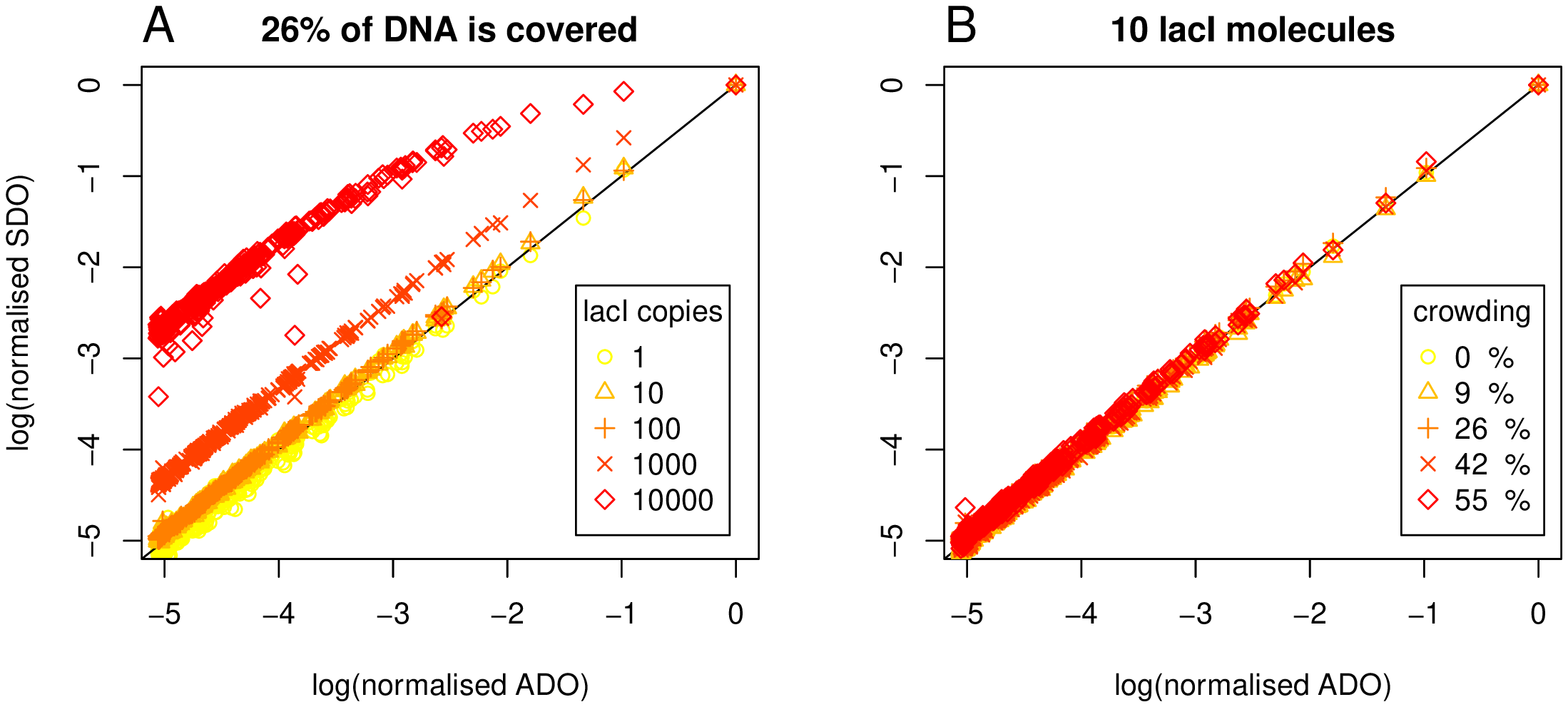}
\end{center}
\caption[ADO and SDO  for various abundances of lacI and crowding on the DNA]{\justifying{\bf ADO and SDO for various abundances of lacI and crowding on the DNA.} We considered the case of the lac repressor TF and $100\ Kbp$ of DNA, which contain the $O_1$ site.  Each system was simulated for $T_l=3000\ s$ (which is the average cell cycle time of \emph{E.coli} \citep{santillan_2004,rosenfeld_2005}) and, for each set of parameters, we considered $X=40$ independent simulations. We considered only the sites that have the binding energy at least $70\%$ of the highest value (the strongest $437$ sites). \onel\ Five different lacI copy numbers: \one $1$, \two $10$, \three $100$, \four $1000$ and \five $10000$. We assumed the case of $3\times 10^4$ copies of non-cognate TFs, which lead to $26\%$ of the DNA being covered. \twol\ Five different non-cognate copy numbers: \one $0$, \two $1\times 10^4$, \three $3\times 10^4$, \four $5\times 10^4$ and \five $7\times 10^4$, and $10$ copies of lacI.}
\label{fig:OccupancyAffinityCrowdingScatterplot}
\end{figure}

In \citep{zabet_2012_model}, we found that, under certain conditions, the occupancy in the simulations cannot always be predicted based on the affinity. To systematically assess the source of the mismatch between affinity derived occupancy (ADO) and simulation derived occupancy (SDO), we considered the case of a bacterial TF (lacI) with biologically plausible parameters and investigated the relationship between affinity and occupancy. Figure \ref{fig:OccupancyAffinityCrowdingScatterplot} contains scatter plots of the SDO vs. ADO at individual sites (at $1\ bp$ resolution) for various crowding levels on the DNA, and various lacI abundances. To eliminate weak sites which will not facilitate the formation of a strong complex with lacI, we recorded only sites with high affinity $E_{\textrm{lacI}}^{j} \geq E_{\textrm{lacI}}^{O_{1}}\times0.7$. We chose this threshold to select the top $0.5\%$ of sites based on the distribution of binding energies, but the value of the threshold can be selected to match any distribution of binding energies.  

Figure \ref{fig:OccupancyAffinityCrowdingScatterplot}\onel\ shows that for $1$ lacI molecule, there is an excellent agreement between ADO and SDO even in the case of crowding on the DNA. The mean ratio of SDO to ADO for $1$ lacl molecule with $26\%$ crowding is $0.966$, within a $95\%$ confidence interval $(0.825, 1.120)$. This suggests that, even in the case of leaky gene expression ($1$ or a few TF molecules), the TF is able to regulate a gene within a cell cycle and the percentage of time the site is occupied is not affected by crowding.

Usually, bacterial TFs number between $10$ and $100$ copies per cell \citep{wunderlich_2009}. In this case, as well as in the case of $1$ lacI molecule, the addition of non-cognate TFs does not appear to introduce a significant difference between ADO and SDO. 

Finally, a few bacterial TFs are known to exist in high copy numbers (e.g. the copy number of CRP is $\approx 1000$ \citep{santillan_2004}) and Figure \ref{fig:OccupancyAffinityCrowdingScatterplot}\onel\ confirms that, in the case of highly abundant bacterial TFs, the ADO diverges from the SDO. In particular, we observed a two-fold increase in SDO, compared to ADO; see Table \ref{tab:resultsCondfidenceIntCopies}. This indicates that certain sites (for example $O_2$, the second strongest site of lacI) will display a higher degree of occupancy than that predicted by affinity.

\begin{table}[!ht]
\centering
\begin{tabular}{| c | c | c | c | c | c |}
\hline
mean &  $0.966$ & $1.081$  & $1.090$ & $1.950$ & $9.782$\\
\hline
lacI copies & $1$ &  $10$ & $100$  & $1000$ & $10000$\\
\hline
$1$ &  & $(0.108, 0.123)$ & $(0.117, 0.131)$ & $(0.973, 0.995)$ & $(8.680, 8.950)$\\
\hline
$10$  & & & $(0.006, 0.012)$ & $(0.860, 0.877)$ & $(8.570, 8.830)$\\
\hline
$100$ &  &  & & $(0.851, 0.868)$ & $(8.560, 8.820)$\\
\hline
$1000$ & & & & & $(7.700, 7.970)$\\
\hline
\end{tabular}
\caption[Confidence intervals around change in ratio SDO/ADO with 26\% crowding.]{\justifying\emph{Confidence intervals around change in ratio SDO/ADO with 26\% crowding.}  $95\%$ t-test confidence interval for the difference in mean ratio SDO/ADO between abundances of lacI transcription factor. For example, moving from $1$ lacI copy to $1000$ copies sees the confidence interval at $(0.880, 0.909)$ - in other words the mean ratio has shifted by nearly $1$. This is reflected in the raw mean values for $1$ copy and $1000$ copies of $1.066$ and $1.960$ respectively.}
\label{tab:resultsCondfidenceIntCopies}
\end{table}

Next, we considered the effect of increased crowding of the DNA by non-cognates on the relationship between ADO and SDO. Figure \ref{fig:OccupancyAffinityCrowdingScatterplot}\twol\ shows that increasing the crowding level has a negligible effect on this relationship and that ADO is a good approximator of SDO at all levels of non-cognate crowding when 10 lacI molecules are modelled; see also Table \ref{tab:resultsCondfidenceIntCrowding}.

\begin{table}[!ht]
\centering
\begin{tabular}{|>{\centering\arraybackslash}m{15mm}|c|c| c | c | c |}
\hline
$\%$ of covered DNA & $0\%$ &  $9\%$ & $26\%$  & $42\%$ & $55\%$\\
\hline
mean & $1.010$ & $0.968$  & $1.081$ & $0.993$ & $1.066$\\
\hline
C.I. &(0.008, 0.012) &  $(-0.035, -0.030)$ & $(-0.076, -0.080)$ & $(-0.011, -0.005)$ & $(0.059, 0.067)$\\
\hline
\end{tabular}
\caption[Effect of crowding on ratio SDO/ADO for 10 lacI molecules.]{\justifying\emph{Effect of crowding on ratio SDO/ADO for 10 lacI molecules.} The table shows the mean SDO/ADO ratio for different levels of crowding. Confidence intervals are from a $95\%$ t-test and show shift in mean ratio from $0\%$ crowding level.}
\label{tab:resultsCondfidenceIntCrowding}
\end{table}

Altogether, non-cognate binding proteins do not affect the occupancy of medium and high affinity sites, in the sense that the SDO of medium and high affinity sites is accurately approximated by the ADO. However, by significantly increasing the abundance of cognate TFs, ADO ceases to be a good approximator of the SDO of medium and high affinity sites. Thus, only cognate abundance influences the occupancy of medium and high affinity sites, while non-cognate TFs have only limited effect. 

The results shown in Figure \ref{fig:OccupancyAffinityCrowdingScatterplot}, use normalised measures of occupancy (ADO and SDO), which are the relative values with respect to the highest rate of occupancy at the strongest site. When analysing the absolute values for occupancy, \citet{wasson_2009} observed that the addition of non-specific DNA binding proteins (nucleosomes in their studies) will reduce the absolute occupancy of cognate TFs. In \sm \ref{seq:appendixSDOabsolute} we show that the SDO increases when the lacI abundance is increased and slightly decreases when the non-cognate abundance is increased, supporting the results from \citep{wasson_2009}.

\subsection{Non-specific sites}

Figure \ref{fig:OccupancyAffinityCrowdingScatterplot} considers only sites with an affinity above a specific threshold. Besides providing more clarity, the rationale for this restriction was twofold: First, there is no clear evidence for the biological relevance of extreme low affinity sites, and second, we are only interested in amounts of occupancy that would be detectable in a biochemical assay (i.e. extreme low affinity binding events are likely not detectable), as the theoretical explanation of observed binding profiles is one of the goals of our research.

\begin{figure}[!ht]
\begin{center}
\includegraphics[scale=0.7,angle=270]{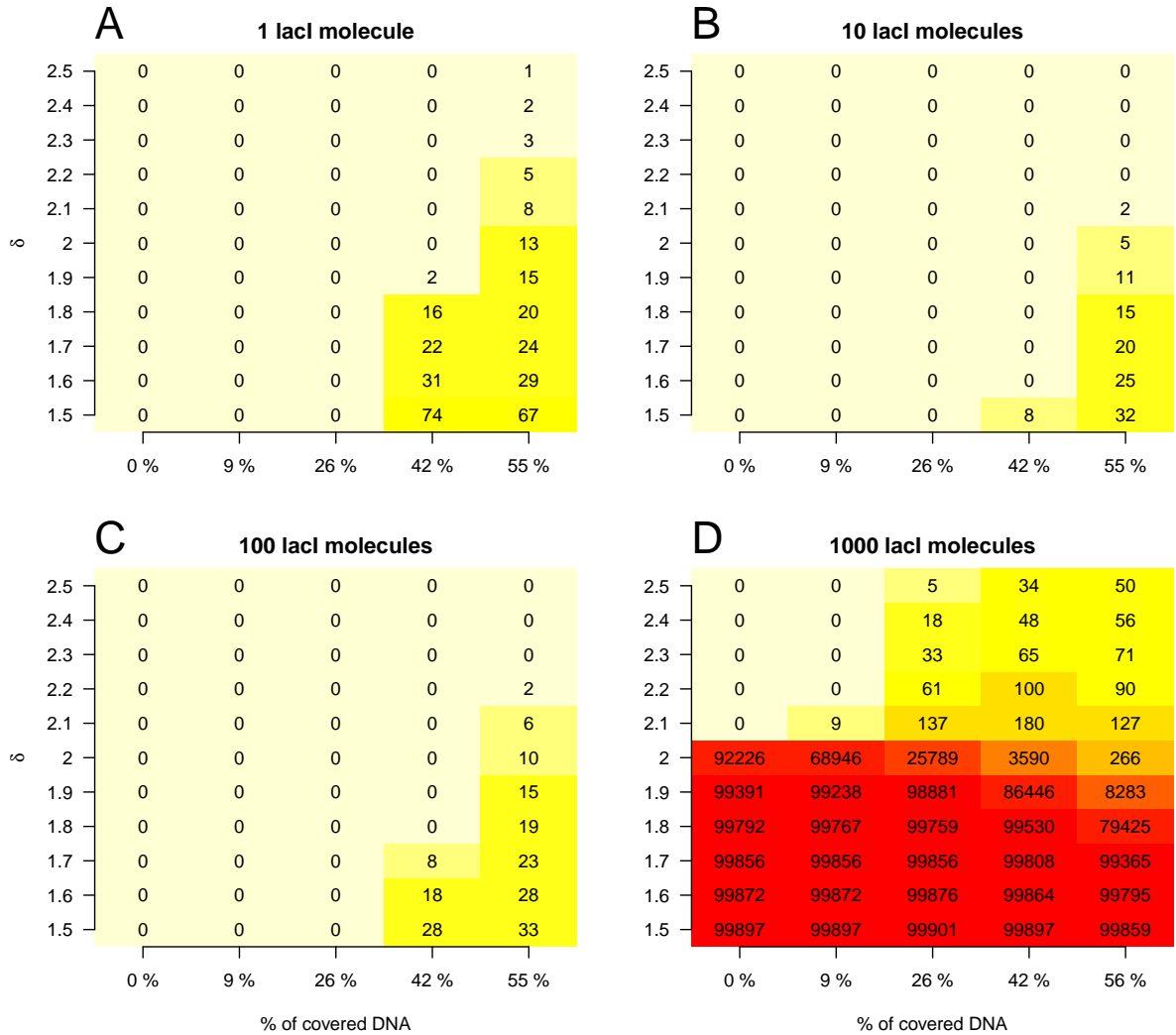}
\end{center}
\caption[Significant deviations between ADO and SDO]{\justifying{\bf Significant deviations between ADO and SDO}. In this graph, we did not consider any affinity cut-off and plotted the number of sites where the ratio between SDO and ADO exceeds  $\delta$ for a range of values of $\delta\in[1.5,2.5]$. There are four cases: \onel\ $1$ lacI molecule, \twol\ $10$ lacI molecules, \threel\ $100$ lacI molecules and \fourl\ $1000$ lacI molecules.}
\label{fig:OccupancyAffinityCrowdingCountsSignificant}
\end{figure}

Figure \ref{fig:OccupancyAffinityCrowdingCountsSignificant} shows heatmaps representing the number of sites where the ratio between SDO and ADO is higher than a factor SDO/ADO $> \delta$. For example, when $\delta>1$, the graph considers the sites where occupancy predicted from affinity underestimates the occupancy observed in the simulations. Interestingly, we did not find any sites where the SDO is lower than the ADO (which we call `false negative' sites), under the various combinations of lacI abundances and crowding levels on the DNA (data not shown).

However, we found sites where SDO $>$ ADO and we call these sites `false positives'. For lacI abundances within [1,100] copies - Figures~\ref{fig:OccupancyAffinityCrowdingCountsSignificant}(\emph{A-C}) - there are tens of sites where the SDO is higher by at least $50\%$ compared to the ADO ($\delta\geq1.5$). These sites appear only for high levels of crowding (at least $42\%$) and their number is increased by increasing the crowding. This means that by increasing the crowding on the DNA the number of sites where SDO is higher than ADO also increases. We also investigated if there is a particular affinity of the sites where the SDO exceeds ADO and found that these sites are usually distributed amongst the medium and non-specific sites; see \sm \ref{seq:appendixSDOADOsignificant}.

When we looked for larger differences between SDO and ADO we saw that by increasing $\delta$ we observed fewer false positive sites. In particular, for  $[1,100]$ copies of lacI, there is no site where the occupancy in the simulations is higher by  $150\%$ (i.e. $\delta \geq 2.5$)  than the value predicted by the affinity. This supports the conclusion from the previous section that the occupancy we observed in the simulations does not significantly deviate from that predicted based on the affinity.

In the case of 1000 copies of lacI, the results differ. Specifically, there appears to be two regimes, namely: \one for $\delta \leq 2$ and \two for $\delta >2$. In the first of these ($\delta\in[1.5,2.0]$), increasing the number of non-cognate molecules reduces the number of sites where the SDO/ADO $< \delta$. In other words, in this regime, increased crowding on the DNA has the opposite effect than that for lower lacI copy numbers (see above): it reduces the number of false positive sites. In the case of $1000$ copies of lacI, the mean SDO/ADO ratio is $\delta_r \approx 2$ (whilst when lacI abundance $\leq$ 100 copies it is approximately $1$) and by adding non-cognates the number of bound cognate molecules at sites whose SDO/ADO $\leq \delta_r $ is reduced (see \sm \ref{seq:appendixBoundlacI}). In turn the mean SDO/ADO ratio will be reduced which in turn explains why the number of false positive sites decreases. In the latter case ($\delta\in(2.0,2.5]$), we observe a similar effect as for lower abundances of lacI, namely that increasing the crowding on the DNA increases the number of bound cognate molecules at sites where SDO/ADO$ > \delta_r$.

\subsection{Considerations on eukaryotic cells}
Eukaryotes typically have $3\times 10^4$ TF copies per cell \citep{biggin_2011}, with some abundances being is high as $3\times 10^6$ copies per cell \citep{biggin_2011}. This higher abundance of TFs comapred to prokaryotes appears to reflect that eukaryotic genomes are much longer, giving much greater space in which TFs can bind \citep{kaplan_2011}. However, at any one time large parts of eukaryotic genome are packed into dense chromatin, and are thus inaccessible to TF binding. For example, in the \emph{D. melanogaster} embryo, on average only $4.1\ Mbp$ of the euchromatic genome of $118\ Mbp$ is  accessible during each early developmental stage \citep{thomas_2011}. This means that, in such eukaryotic cells, we have accessible DNA that is similar in length to that considered in this study (the \emph{E.coli} genome is approximately $4.6\ Mbp$), but with TFs in much greater abundance. This begs the question of whether the relationship between occupancy and affinity that we observe when simulating the prokayrotic case (lacI around the $O_1$ site) is still true in the context of eukaryotic systems with TFs that have $\sim 10^4$ copies or more.

It is clear from Figure \ref{fig:OccupancyAffinityCrowdingScatterplot} that increasing the abundance of cognate TFs up to $10^4$, increases the number of medium affinity sites that display significantly higher occupancy; see also Table \ref{tab:resultsCondfidenceIntCopies}.  This observation remains true for different levels of crowding on the DNA as introduced by the presence of non-cognate TFs (no crowding, low crowding and medium crowding) (data not shown). Furthermore, at such high levels of cognate abundance almost all sites display a much higher occupancy than that predicted from their affinity. For example, the occupancy of the second strongest site of lacI ($O_2$) becomes approximately equal to that of the strongest one ($O_1$), although there is a  large difference in affinity between the two sites. This observation suggests that high TF abundance makes strong and weak sites less distinguishable, which would hinder a quantitative readout for the regulation of gene expression in the cell. 

\begin{figure}[!ht]
\begin{center}
\includegraphics[scale=0.7,angle=270]{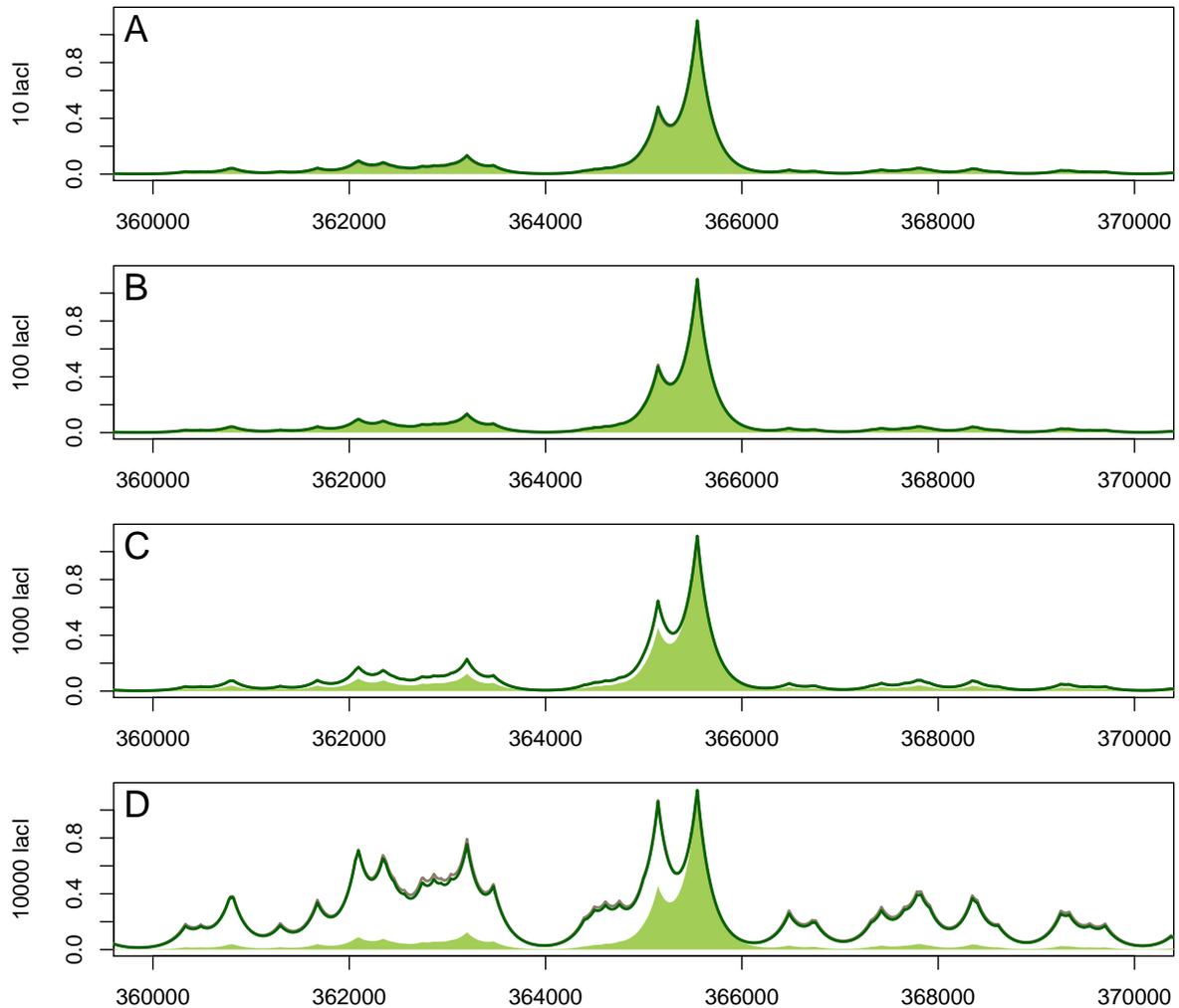}
\end{center}
\caption[SDO and ADO landscape for various cognate and non-cognate abundances]{\justifying{\bf SDO and ADO landscape for various cognate and non-cognate abundances.} We considered the case of the lac repressor TF and $100\ Kbp$ of DNA, which contain the $O_1$ site. In each chart the solid green line is the SDO at one of four levels of lacI abundance, and the filled green region is the ADO. The SDO shown is calculated with 0 non-cognate molecules; calculations for 10\% and 26\% non-cognate abundance show no visible deviation from the 0 non-cognate case (hence not shown). The SDO was calculated at four lacI abundances: \onel\ $10$, \twol\ $100$, \threel\ $1000$ and \fourl\ $10000$ molecules. Each system was simulated for $T_l=3000\ s$ and for each set of parameters we consider $X=40$ independent simulations. We considered only the sites that have the binding energy at least $70\%$ of the highest value (the strongest $437$ sites). We converted the single nucleotide resolution into expected ChIP-seq profiles as proposed in \citep{kaplan_2011}; see \sm \ref{seq:appendixChIPprofile}.}
\label{fig:OccupancyAffinityLandscape}
\end{figure}

Above, we considered occupancy and affinity at single nucleotide resolution. Figure \ref{fig:OccupancyAffinityLandscape} shows a theoretical TF binding profile over a locus of the \emph{E.coli} genome as calculated using GRiP, demonstrating the progressive effect on occupancy of increasing TF abundance. (The theoretical profiles are generated using a method described by \citet{kaplan_2011} for modelling ChIP-seq profiles; see \sm \ref{seq:appendixChIPprofile}).  Each chart plots the ADO and SDO, and shows that for low copy numbers ($[10,100]$ copies per cell), the profile of the ADO (filled region) matches the profile of SDO (solid line) with high accuracy for the cases of no crowding on the DNA ($0$ non-cognate molecules) and medium crowding on the DNA ($3\times 10^4$ non-cognate molecules). This would imply that, in bacterial cells (i.e. when TF abundance is relatively low), the binding of TFs to their target sites is not affected by competition with other molecules, and occupancy is predominantly a factor of, and is accurately modeled by, affinity. However, when TFs are highly abundant ($[10^3,10^4]$ copies per cell), as is common in eukaryotic systems, the level of affinity is not the sole determinant of occupancy on the DNA. In other words, the amount of time spent bound is determined not just by the encoded information in the DNA (nucleotide composition of binding sites) and DNA accessibility, but by the abundance of TFs in the system (mainly cognate TF abundance, but small effects from non-cognates were observed).

\begin{figure}[!ht]
\begin{center}
\includegraphics[scale=0.8,angle=270]{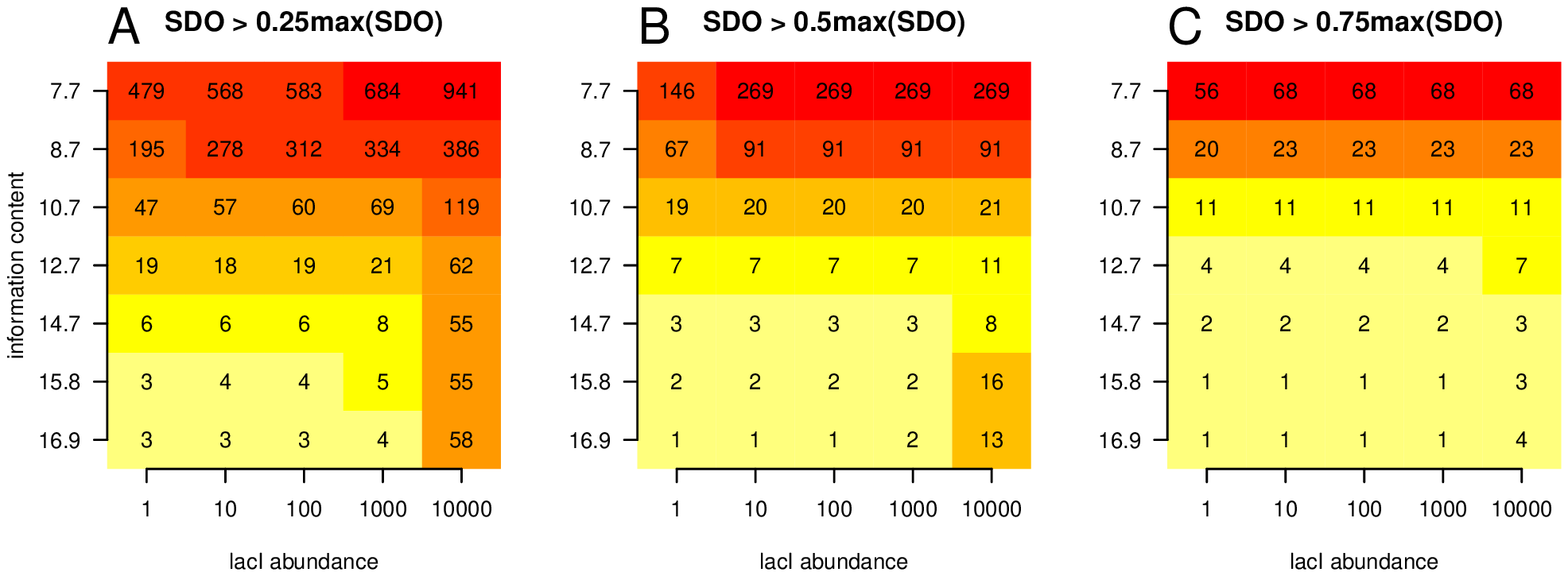}
\end{center}
\caption[The relationship between information content of the PWM motif and the abundance of TF]{\justifying{\bf The relationship between information content of the PWM motif and the abundance of TF}. This graph represents the number of sites that display an occupancy in the simulation that is higher than the following thresholds: \onel\ $0.25\cdot \max\left(SDO\right)$,  \twol\ $0.50\cdot \max\left(SDO\right)$ and  \threel\ $0.75\cdot \max\left(SDO\right)$. There were no non-cognate TFs in these cases and occupancy was calculated at abundances of lacI $\in \{1,10,100,1000,10000\}$. Information content of the lacI motif was reduced by succesively removing the rightmost column of the PWM (see \sm \ref{seq:appendixLowerInfo}). In general the number of high occupancy sites is increased by both increased lacI abundance (compare the values on each row) and reduced information content (compare the values on each column). In \twol\, at the highest lacI abundance, there are several cases where the number of highly occupied sites decreases with reducing the information content (from 16 to 8) contrary to the pattern at other abundances and/or thresholds. This can be explained by the fact that, in order to reduce the information content, we removed certain base pairs from the lacI motif, which can introduce biases in the affinity landscape. These biases can lead to small deviations from the expected results, particularly, in the cases where there are few sites and the TF has high abundance. For example, in the case of the $10000$ copies of lacI with the full motif, there are sites that display an occupancy of $0.6\cdot \max\left(SDO\right)$, while, in the case of $10000$ copies of lacI with information content $14.7$, those sites will display an occupancy of $0.4\cdot \max\left(SDO\right)$.}
\label{fig:OccupancyAffinityAbundanceInformationSignificant}
\end{figure}

Finally, bacterial TFs have PWMs with higher information content compared to the eukaryotic TFs \citep{stormo_1998,wunderlich_2009}, e.g., for  lacI, $I_{lacI}=16.9\ bits$. To investigate the influence of information content on the number of highly occupied sites observed in the simulations, we removed  positions from the end of the lacI motif and performed the simulations at various abundances of lacI on naked DNA (i.e. no non-cognate TF molecules). In total, we considered six cases, which resulted in the information content of the reduced lacI motif being: \one $I_{lacI_1}=15.8$, \two $I_{lacI_2}=14.7$, \three $I_{lacI_3}12.7$, \four $I_{lacI_4} = 10.7$, \five $I_{lacI_5} = 8.7$ and \six  $I_{lacI_6} = 7.7$; see \sm \ref{seq:appendixLowerInfo}. Figure \ref{fig:OccupancyAffinityAbundanceInformationSignificant} shows that, by selecting an arbitrary threshold (certain percent of the highest value of SDO), the number of sites with SDO higher than the threshold increases both as the abundance of lacI increases (compare the values on each row in Figure \ref{fig:OccupancyAffinityAbundanceInformationSignificant}), and as the information content of the motif decreases (compare the values on each column in Figure \ref{fig:OccupancyAffinityAbundanceInformationSignificant}). Note that the former (the dependence of the SDO on the TF abundance) was already shown in Figure \ref{fig:OccupancyAffinityCrowdingScatterplot} and Figure \ref{fig:OccupancyAffinityLandscape}. Hence, in eukaryotic systems, we can expect a two fold increase in the number of sites with high SDO from both the greater TF abundance \citep{biggin_2011} \emph{and} from the likely lower information content of the average eukaryotic PWM \citep{wunderlich_2009}.

Note that by removing certain positions from the end of the lacI motif, we reduced the information content in a biased way and this can lead to small variations in the occupancy, particularly, in the case when there are a few sites that display high occupancy. Nevertheless, this approach to change the information content does not influence the general result, that TFs with lower information content motifs display more dramatic change in the number of sites highly occupied compared to TFs with higher information content motifs.

\section{Discussion}
Transcription factors perform a combination of three-dimensional diffusion and one-dimensional random walk on the DNA when they search for their target sites. Inherently, this mechanism leads to the binding of TFs not only to their target sites, but also to other, lower affinity sites on the DNA. In this context, it becomes important to understand the relationship between affinity (how strongly a TF binds to a site on the DNA) and occupancy (the residence time of a TF on a site).

Often it is assumed that the relative occupancy of a TF measured experimentally (say, in a ChIP assay) is indicative of the relative affinity, and many studies infer a TF's affinity by de novo motif analysis based on the most highly occupied sites (those showing the strongest ChIP enrichment). This assumption is flawed when there is divergence between occupancy and affinity for these highly occupied sites. Although this approximation proved to have good accuracy in the inference of position weight matrices in many cases \citep[e.g.]{adryan_2007b}, there are also examples where the method seems to fail \cite[e.g.]{zeitlinger_2007}. These cases refer to situations where false positive prediction (sites that have low affinity but display high occupancy) or false negative prediction (sites that have high affinity but display low occupancy) could have influenced the success of the study.

Our results indicate that by adding non-cognate TFs, the absolute occupancy of binding sites by cognate TF molecules is reduced (see \sm \ref{seq:appendixSDOabsolute}). The reduction in the absolute value of the occupancy is a consequence of the competition of TFs for the limited amount of DNA. \citet{wasson_2009} observed the same effect, although they used a different approach (a statistical thermodynamics model) to estimate the occupancy. However, in their study, they did not look at the occupancy relative to the highest value (the quantitative readout of binding events).  

We found that the abundance of non-cognate TFs has a limited effect on the normalised occupancy of low, medium and high affinity sites; see Figure \ref{fig:OccupancyAffinityCrowdingScatterplot}\twol\ and Figure \ref{fig:OccupancyAffinityCrowdingCountsSignificant}. Nevertheless, there are several sites (in the order of tens), where the addition of non-cognate TFs leads to significant deviations of the observed occupancy derived from simulation (SDO) from that derived from affinity (ADO). This result is supported by recent experimental evidence, where the authors showed that lac repressor occupancy increases at lower sites (far away from the $O_1$ site), when the crowding in the cell increases (and, thus, the crowding on the DNA increases as well) \citep{kuhlman_2012}.

Bacterial TFs are expressed at low copy numbers (between $10$ and $100$) \citep{wunderlich_2009} and they have only a few strong sites that are highly specific  \citep{stormo_1998,wunderlich_2009}. This suggests that, in the case of bacterial gene regulation, affinity controls the relative occupancy of the specific sites (acting as a local fine tuning mechanism), while the crowding level on the DNA controls the global occupancy of the sites (acting as a global regulator).

We also investigated under which conditions the normalised occupancy of the medium and high affinity sites is affected. Our results confirmed that for TFs with $10^3-10^4$ copies per cell and approximately $4\ Mbp$ of available DNA, the occupancy is higher than that predicted by affinity, irrespective of the abundance of non-cognate TFs. Eukaryotic systems have TFs with high  abundance (on average $3\times 10^4$ copies per cell) \citep{biggin_2011} and although they have  much larger genomes, only a small proportion of this is accessible to TFs (e.g., $\approx 4\ Mbp$ in early developmental stages of \emph{D. melanogaster}) \citep{thomas_2011}. This suggests that the rate of false positive binding events (higher occupancy than predicted by affinity) is significant in eukaryotic cells; see Figure \ref{fig:OccupancyAffinityLandscape}. Note that our model is applicable only to TFs residing in the nucleoplasm and, thus, when we mention TF abundance in eukaryotic systems we refer to nuclear abundance of TFs \citep{fowlkes_2008}. 

\citet{kaplan_2011} investigated the relationship between experimentally measured occupancy (from ChIP-seq experiments) and that  predicted using a  hidden Markov model, and found that the highest correlation between the two was on average $\sim 0.7$. To achieve this correlation they assumed real TF abundances that were previously measured in \emph{D. melanogaster} nuclei \citep{fowlkes_2008}, but they did not adapt the abundances of TFs to the size of the analysed DNA segment. In \citep{zabet_2012_subsystem}, we showed that, when the number of bound TF molecules is not changed in such a subsystem (a simulated entity smaller than the genome), the correlation coefficient between the occupancy of the full system and the occupancy of the subsystems can be as low as $0.4$. This result is also shown in Figures \ref{fig:OccupancyAffinityCrowdingScatterplot} and \ref{fig:OccupancyAffinityLandscape}, which confirm that an increase in cognate TF copy number can lead to a reduction in the correlation between occupancy and affinity landscape. Thus, one method to increase the correlation between the predicted and observed occupancy consists of adapting the abundance levels of the TFs with one of the methods presented in \citep{zabet_2012_subsystem}.

In addition, this higher number of highly occupied sites is also influenced by the information content of the motif. In Figure \ref{fig:OccupancyAffinityAbundanceInformationSignificant}, we showed that, by reducing the information content, the number of sites with high SDO increases, but also that the effects of the increase in TF abundance on the highly occupied sites is more dramatic. In other words, by increasing the abundance of a TF with a PWM with lower information content, we observed a larger increase in the number of highly occupied sites compared to the case of a TF with a PWM with higher information content; compare different rows in Figure \ref{fig:OccupancyAffinityAbundanceInformationSignificant}. This suggests that, in the case of eukaryotic systems (which have TFs with lower information content PWMs \citep{wunderlich_2009} and higher abundances \citep{biggin_2011}), the effects of TF abundance on the number of `false positive' sites is more severe than in the case of bacterial cells.

Our approach to reduce the information content (by removing positions from the end of the lacI motif) is prone to introduce biases in the results, in particular, at high abundance of the TF and low number of highly occupied sites; see Figure \ref{fig:OccupancyAffinityAbundanceInformationSignificant}\twol. A different approach to reduce the information content could be to add non-specific sites uniformly when constructing the PWM, but we anticipate this would lead to similar results, namely: in the case of lower information content motifs, a change in the abundance of TF has more drastic effects on the number of highly occupied sites, compared to the case of higher information content motifs. Nevertheless,  the details of this applying a different approach  to reduce the information content need to be left for further research as it is beyond the scope of this manuscript.

Finally, we found that the increase in occupancy caused by the addition of cognate molecules can be reduced by adding non-cognate molecules. Figure \ref{fig:OccupancyAffinityCrowdingCountsSignificant}\fourl\ shows that while, in the case of empty DNA, most of the sites display an occupancy in the simulations that is higher by at least $100\%$ than that predicted from affinity; in the case of high crowding on the DNA, only several hundred sites display such a difference between SDO and ADO. However, this difference is still large, in the order of $70\%$.

\section{Acknowledgments}
We would like to thank Mark Calleja for his support with configuring our simulations to run on CamGrid and  Robert Stojnic for useful discussions and comments.  This work was supported by the Medical Research Council [G1002110 to N.R.Z.]. R.F. is supported by a BBSRC studentship and B.A. is a Royal Society University Research Fellow. 

\appendix
    \begin{center}
      {\bf APPENDIX}
    \end{center}

\section{TF parameters}\label{seq:appendixTFparams}

The default parameters used here were previously derived in \citep{zabet_2012_model} and \citep{zabet_2012_subsystem} and are listed in Table \ref{tab:modelTFparams}. 

\begin{table}
\centering
\begin{tabular}{| p{5cm} | r | r | l |}
\hline
\textbf{parameter} & \textbf{lacI} & \textbf{non-cognate}&  \textbf{notation}\\ \hline
copy number &  \multicolumn{2}{|c|}{see main manuscript} & $TF_{x}$\\ \hline
motif sequence &  see Table \ref{tab:lacI3gPWM} & - &\\ \hline
energetic penalty for mismatch &  $1\ K_BT$ & $13\ K_BT$ & $\varepsilon^{*}_{x}$\\ \hline
nucleotides covered on left &  $0\ bp$ & $23\ bp$ & $TF^{\textrm{left}}_{x}$\\ \hline
nucleotides covered on right &  $0\ bp$ & $23\ bp$ & $TF^{\textrm{right}}_{x}$\\ \hline
association rate to the DNA &    \multicolumn{2}{|c|}{see main manuscript}  & $k^{\textrm{assoc}}_{x}$\\ \hline
unbinding probability &  $0.001474111$ & $0.001474111$ & $P^{\textrm{unbind}}_{x}$\\ \hline
probability to slide left &  $0.4992629$ & $0.4992629$ & $P^{\textrm{left}}_{x}$\\ \hline
probability to slide right &  $0.4992629$ & $0.4992629$ & $P^{\textrm{right}}_{x}$\\ \hline
probability to dissociate completely when unbinding &  $0.1675$ & $0.1675$ & $P^{\textrm{jump}}_{x}$\\ \hline
time bound at the target site &  $1.18E-6\ s$ & $ 0.3314193\ s$ & $\tau^{0}_{x}$\\ \hline
the size of a step to left &  $1\ bp$ & $1\ bp$ & \\ \hline
the size of a step to right &  $1\ bp$ & $1\ bp$ & \\ \hline
variance of repositioning distance after a hop &  $1\ bp$ & $1\ bp$ & $\sigma^{2}_{\textrm{hop}}$\\ \hline
the distance over which a hop becomes a jump  &  $100\ bp$ & $100\ bp$ & $d_{\textrm{jump}}$\\  \hline
the proportion of prebound molecules & $0.0$ &  $0.9$ &\\  \hline
affinity landscape roughness & - & $1.0\ K_BT$ &\\ 
\hline
\end{tabular}
\caption{\emph{TF species default parameters}}
\label{tab:modelTFparams}
\end{table}

The PWM of lacI was presented in \citep{zabet_2012_subsystem} and is also listed in Table \ref{tab:lacI3gPWM}.

\begin{table*}
\begin{center}
  \begin{tabular}{ |r | r | r | r | r |}
    \hline
     &   \multicolumn{4}{|c|}{\textbf{PWM}}  \\ \hline
    \textbf{Position} & A  &  C & G & T  \\ \hline
    $1$ & $0.6200$ & $-0.6900$ & $0.1400$ & $-0.6900$\\ \hline
    $2$ & $0.6200$ & $-0.6900$ & $0.1400$ & $-0.6900$\\ \hline
    $3$ & $0.1600$ & $0.1400$ & $-0.6900$ & $0.1800$\\ \hline
    $4$ & $0.1600$ & $-0.6900$ & $-0.6900$ & $0.6200$\\ \hline
    $5$ & $-0.7000$ & $-0.7000$ & $0.9000$ & $-0.7000$\\ \hline
    $6$ & $-0.6900$ & $-0.6900$ & $-0.6900$ & $0.9300$\\ \hline
    $7$ & $0.0077$ & $-0.0084$ & $-0.0073$ & $0.0083$\\ \hline
    $8$ & $0.0077$ & $-0.0084$ & $-0.0073$ & $0.0083$\\ \hline
    $9$  & $0.0077$ & $-0.0084$ & $-0.0073$ & $0.0083$\\ \hline
    $10$ & $0.0077$ & $-0.0084$ & $-0.0073$ & $0.0083$\\ \hline
    $11$ & $0.0077$ & $-0.0084$ & $-0.0073$ & $0.0083$\\ \hline
    $12$ & $0.0077$ & $-0.0084$ & $-0.0073$ & $0.0083$\\ \hline
    $13$ & $0.0077$ & $-0.0084$ & $-0.0073$ & $0.0083$\\ \hline
    $14$ & $0.0077$ & $-0.0084$ & $-0.0073$ & $0.0083$\\ \hline
    $15$ & $0.0077$ & $-0.0084$ & $-0.0073$ & $0.0083$\\ \hline
    $16$ & $0.6200$ & $-0.6900$ & $0.1400$ & $-0.6900$\\ \hline
    $17$ & $-0.7000$ & $0.9000$ & $-0.7000$ & $-0.7000$\\ \hline
    $18$ & $0.9300$ & $-0.6900$ & $-0.6900$ & $-0.6900$\\ \hline
    $19$ & $0.9300$ & $-0.6900$ & $-0.6900$ & $-0.6900$\\ \hline
    $20$ & $-0.6900$ & $0.1400$ & $-0.6900$ & $0.6200$\\ \hline
    $21$ & $-0.6900$ & $0.1400$ & $-0.6900$ & $0.6200$\\ \hline
  \end{tabular}
\end{center}
\caption{lacI PWM} \label{tab:lacI3gPWM}
\end{table*}

\section{Measuring the occupancy in the simulations} \label{seq:appendixErgodicity}

\begin{figure}[htp]
  	\centering
	\includegraphics[scale=0.7]{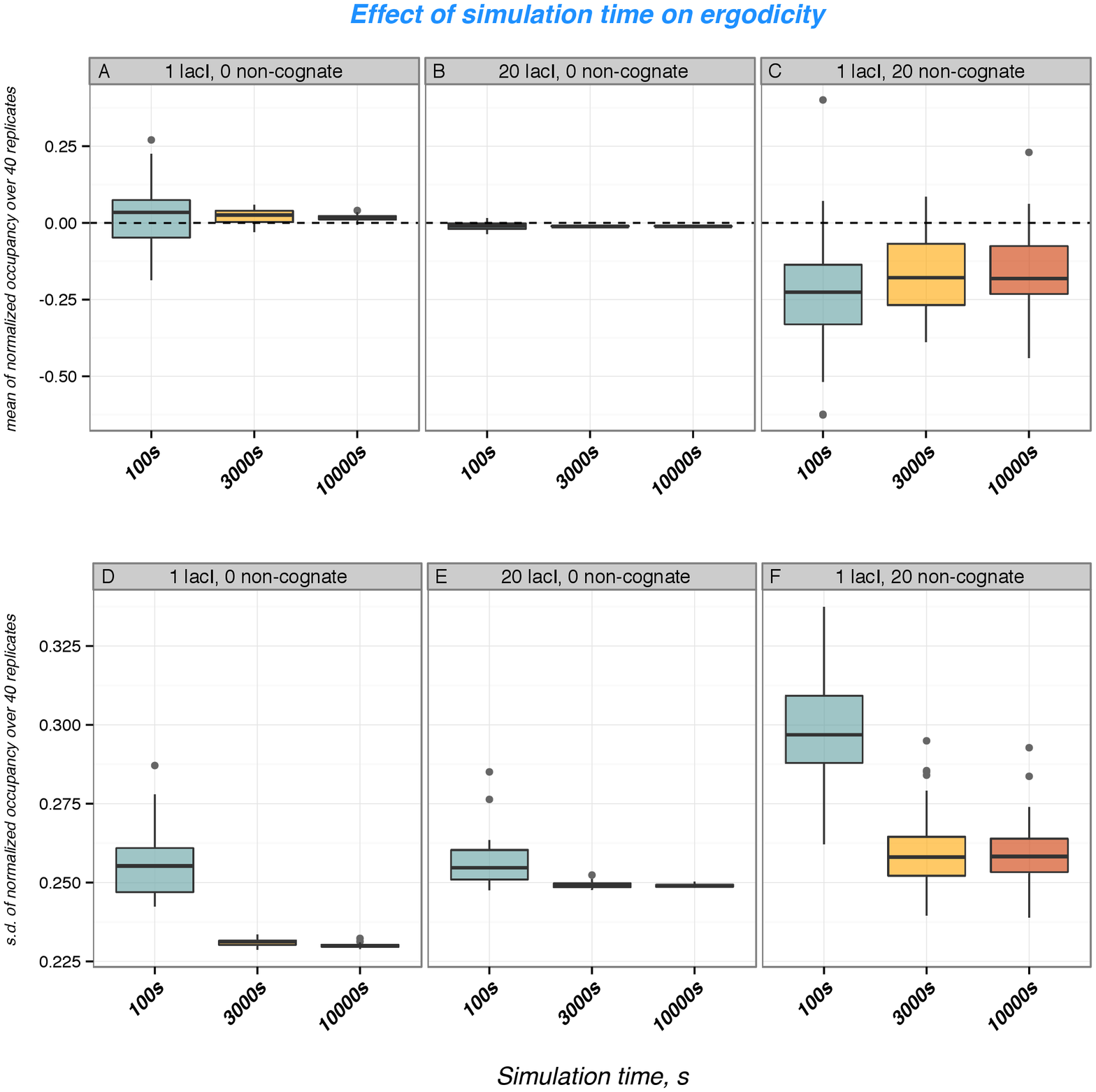}
\caption[Comparing the time average to the ensemble average for various abundances of cognate and non-cognate molecules]{\justifying\emph{Comparing the time average to the ensemble average for various abundances of cognate and non-cognate molecules} The system consists of $1\ Kbp$ of DNA which contains the $O_1$ site. There are three cases with respect to the amounts of TFs: \one $1$ lacI molecule and $0$ non-cognates, \two $20$ lacI molecules and $0$ non-cognates and \three $1$ lacI molecules and $20$ non-cognates. In addition, we considered three values for the simulation time when computing the time and hybrid averages: \one $T_l=100\ s$,  \two $T_l=3000\ s$ and  \three $T_l=10000\ s$. \onel, \twol\ and \threel the boxplots represent the mean of the logarithm of the ratio between the time average and the ensemble average over $40$ replicates. A value of $0$ indicates that the time average is equal to the ensemble average. \fourl, \fivel\ and \sixl the boxplots represent the standard deviation of the logarithm of the ratio between the time average and the ensemble average over $40$ replicates. The sites that have a binding energy lower than $30\%$ of the highest value ($423$) sites were removed. By increasing the simulation time, both the mean and the standard deviation of the logarithm of the ratio between the time average and the ensemble average  tend to $0$, showing that a longer simulation time leads to smaller differences between time and ensemble averages.}\label{fig:ergodicityTimeEnsembleComparison}
\end{figure}

Figure \ref{fig:ergodicityTimeEnsembleComparison} plots the distribution of the logarithm of the ratio between the time and the ensemble averages for the strongest $577$ sites. One can observe that by increasing the simulation time, bot the time average and the hybrid average will deviate from the ensemble average. Furthermore, Figure \ref{fig:ergodicityTimeEnsembleComparison} confirms that the hybrid average performed using $40$ independent replicates, each simulated for $3000\ s$ is a good estimate for the ensemble average. 

\section{System size reduction accuracy} \label{seq:appendixSubsystem}

\begin{figure}[htp]
  \begin{center}
  		\includegraphics[scale=0.9, angle=270]{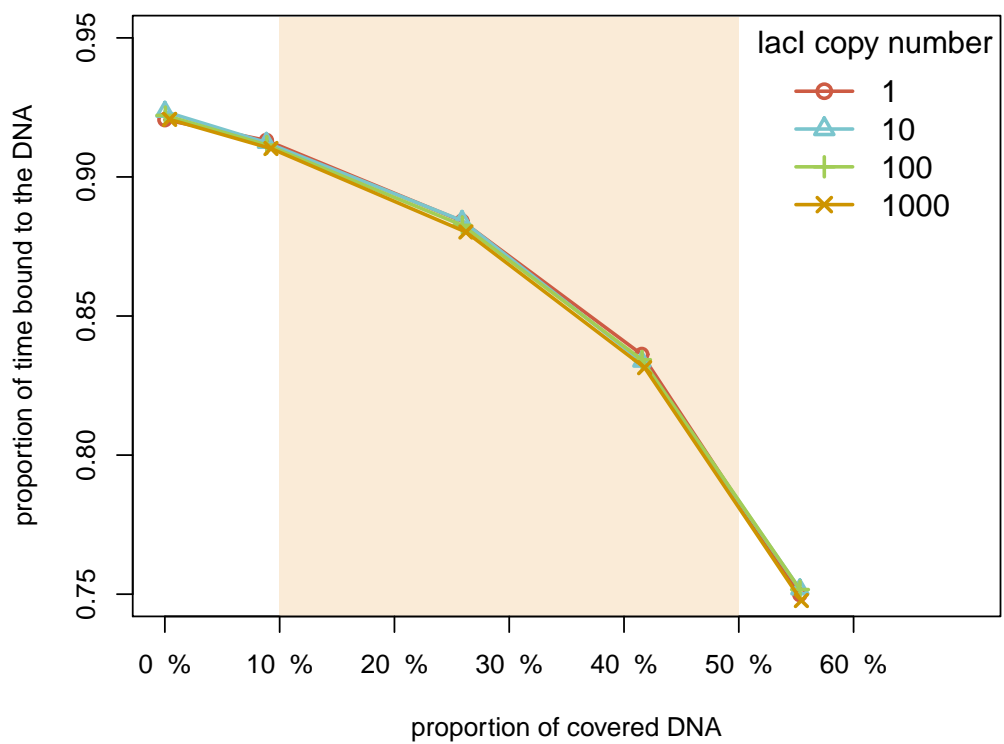}
  \end{center}
\caption[The proportion of time the lacI molecules spend bound to the DNA in the full system, when the crowding on the DNA is altered by changing the abundance and association rate of non-cognate TFs]{\justifying\emph{The proportion of time the lacI molecules spend bound to the DNA in the full system, when the crowding on the DNA is altered by changing the abundance and association rate of non-cognate TFs}.  We performed a set of $20$ simulations of the full system each lasting:  \one $3\ s$ for $1$ lacI,  \two $2\ s$ for $10$ lacI,  \three $1\ s$  for $100$ lacI and  \four $1\ s$  for $1000$ lacI. The shaded area indicates values that are biological plausible.}\label{fig:methodsPropTimeDNAlacI}
\end{figure}

The association rate model required the determination of the actual time spent on the DNA. The proportion of time the lacI molecules spend on the DNA varied if the association rate was fixed to $k^{\textrm{assoc}}_{\textrm{lacI}}=2400\ s^{-1}$, while the percentage of the covered DNA was raised by increasing both the abundance and association rate of non-cognate TFs. The values of the proportion of time the lacI molecules spend on the DNA are plotted in Figure \ref{fig:methodsPropTimeDNAlacI}

\begin{figure}[htp]
  \begin{center}
  		\includegraphics[scale=0.75, angle=270]{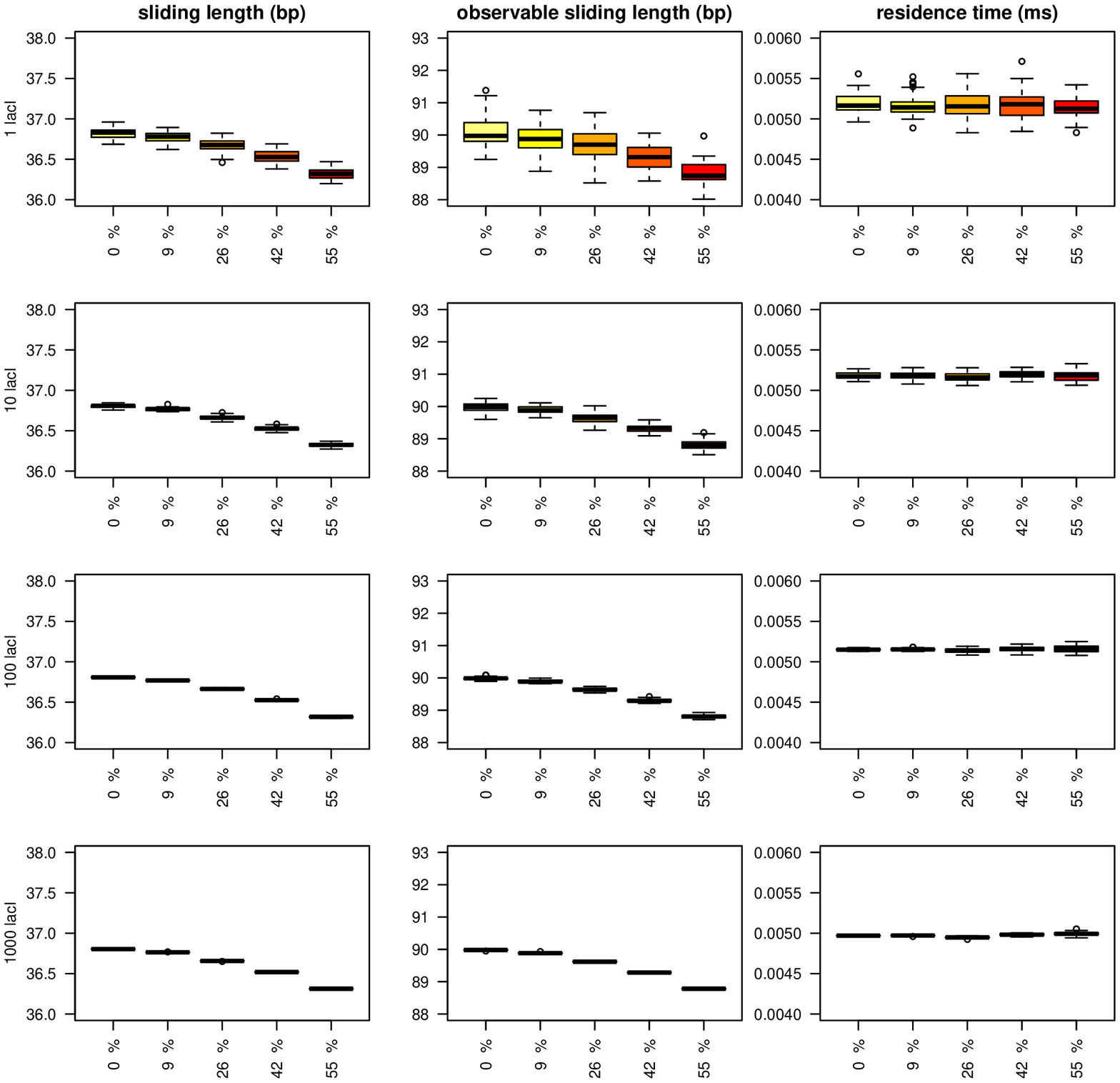}
  \end{center}
\caption[One dimensional statistics for various levels of non-cognate TFs]{\justifying\emph{One dimensional statistics for various levels of non-cognate TFs}.  We performed a set of $X=20$ simulations of the  $100\ Kbp$ subsystem each lasting $T_l=3000\ s$, using the parameters presented in the main manuscript and the parameters from Table \ref{tab:modelTFparams}.}\label{fig:methods1Dstatistics}
\end{figure}

Furthermore, it is important to test whether the one dimensional statistics (sliding length and residence time) are affected by increasing the number of non-cognate TFs. Figure \ref{fig:methods1Dstatistics} shows that, for biologically plausible values, for the proportion of covered DNA (between $10\%$ and $50\%$), the sliding length and the residence time deviate only negligibly from the values that were estimated previously \citep{zabet_2012_model}.

\section{The dependence of absolute occupancy on TF competition} \label{seq:appendixSDOabsolute}

\begin{figure}[htp]
\begin{center}
\includegraphics[scale=0.7,angle=270]{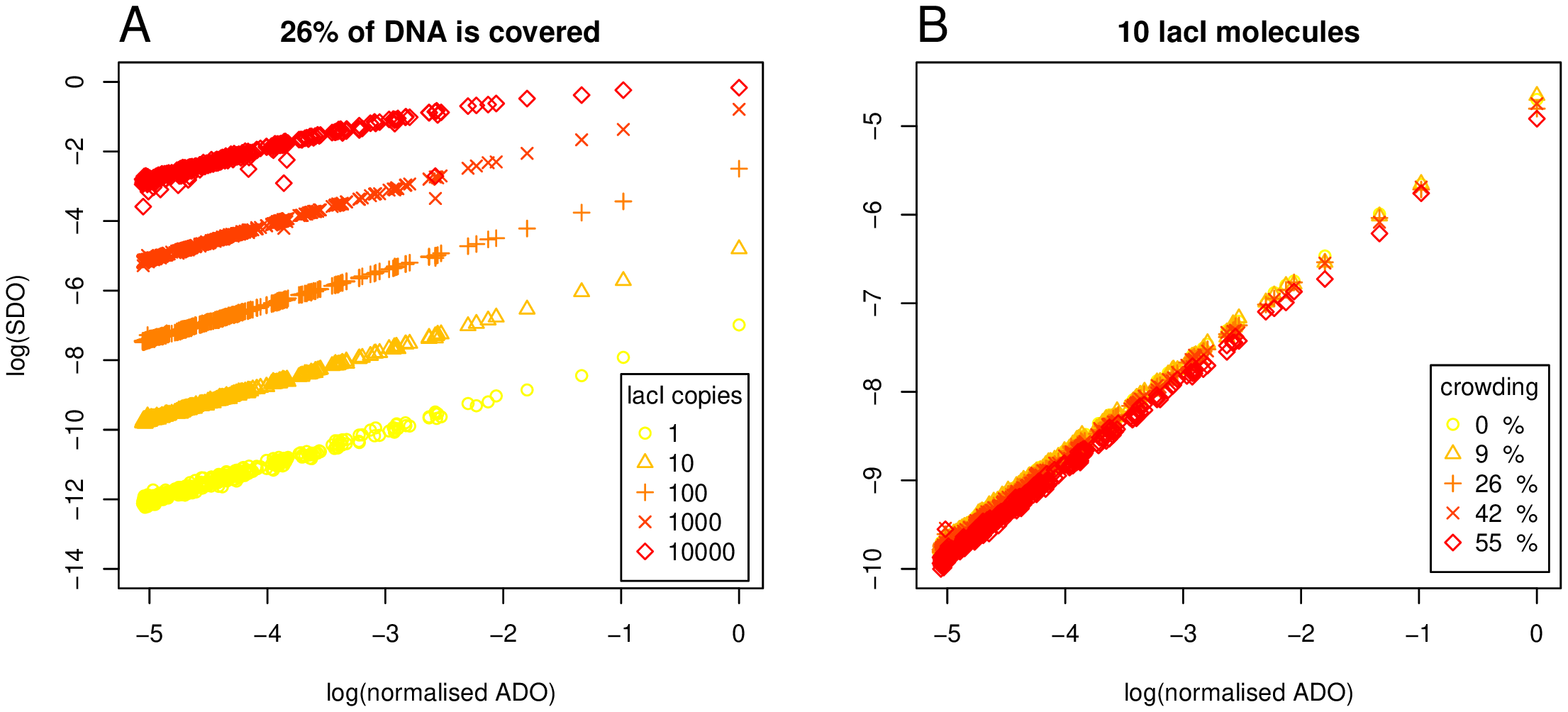}
\end{center}
\caption[ADO and SDO  for various abundances of lacI and crowding on the DNA]{\justifying{\bf ADO and SDO for various abundances of lacI and crowding on the DNA.} This is the same as Figure 3 in the main manuscript, except that the SDO was not normalised.}
\label{fig:OccupancyMaximum}
\end{figure}

Figure \ref{fig:OccupancyMaximum} shows that the absolute SDO (not normalised to the maximum value) is not significantly affected by crowding on the DNA, but strongly depends on the abundance of lacI molecules.

\section{The average number of bound lacI molecules} \label{seq:appendixBoundlacI}

\begin{figure}[htp]
  \begin{center}
  		\includegraphics[scale=0.75, angle=270]{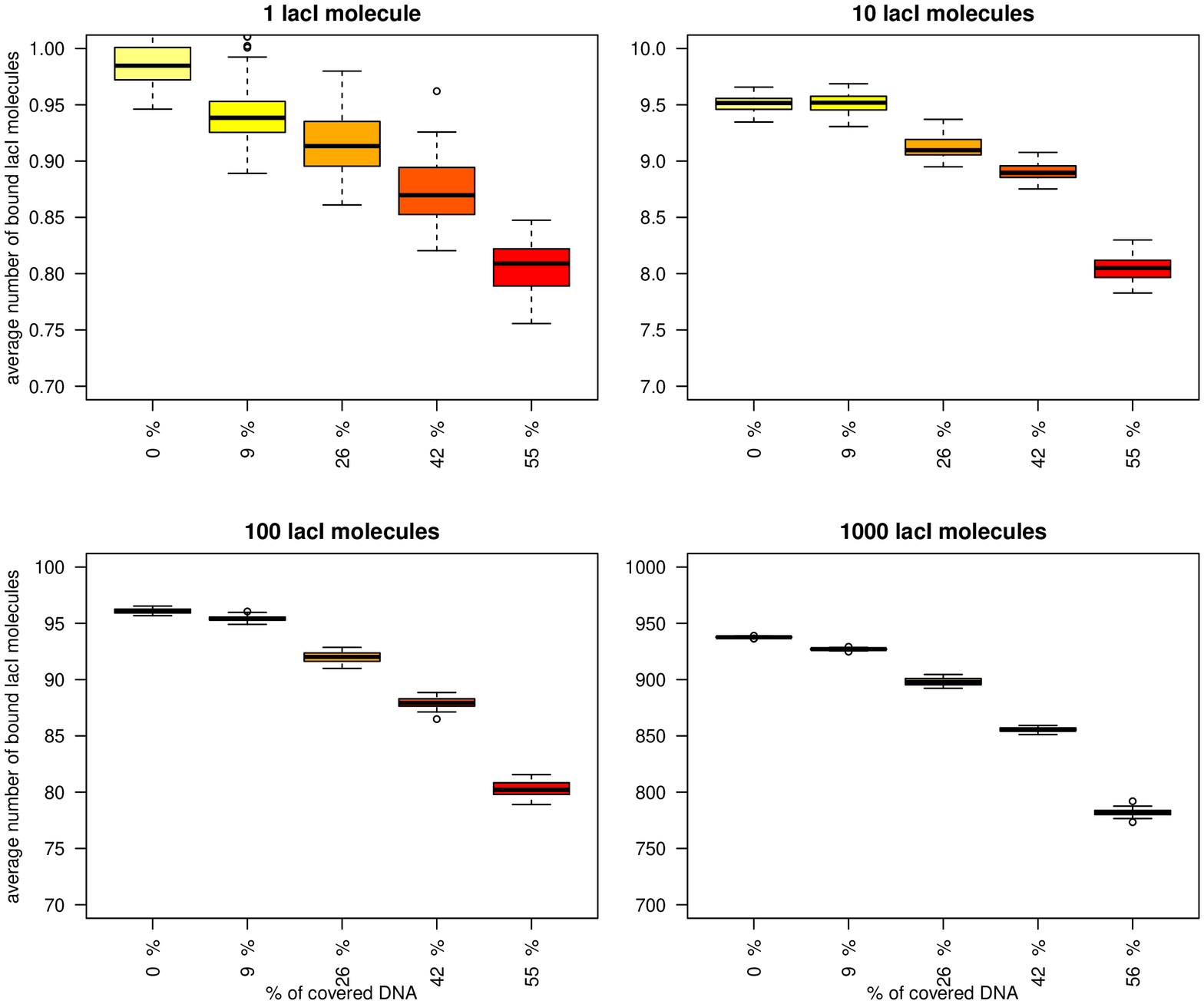}
  \end{center}
\caption[The average number of bound molecules for various crowding levels and various lacI abundances]{\justifying\emph{The average number of bound molecules for various crowding levels and various lacI abundances}.  We performed a set of $X=40$ simulations of the  $100\ Kbp$ subsystem each lasting $T_l=3000\ s$, using the parameters presented in the main manuscript and the parameters from Table \ref{tab:modelTFparams}.}\label{fig:boundlacI}
\end{figure}

Figure \ref{fig:boundlacI} confirms that there is a reduction in the number of bound lacI molecules when the crowding on the DNA is increased by adding more non-cognate molecules. This is valid for all lacI abundances.

\section{Significant difference between  SDO and ADO} \label{seq:appendixSDOADOsignificant}

\begin{figure}[htp]
  \begin{center}
  		\includegraphics[scale=0.7, angle=270]{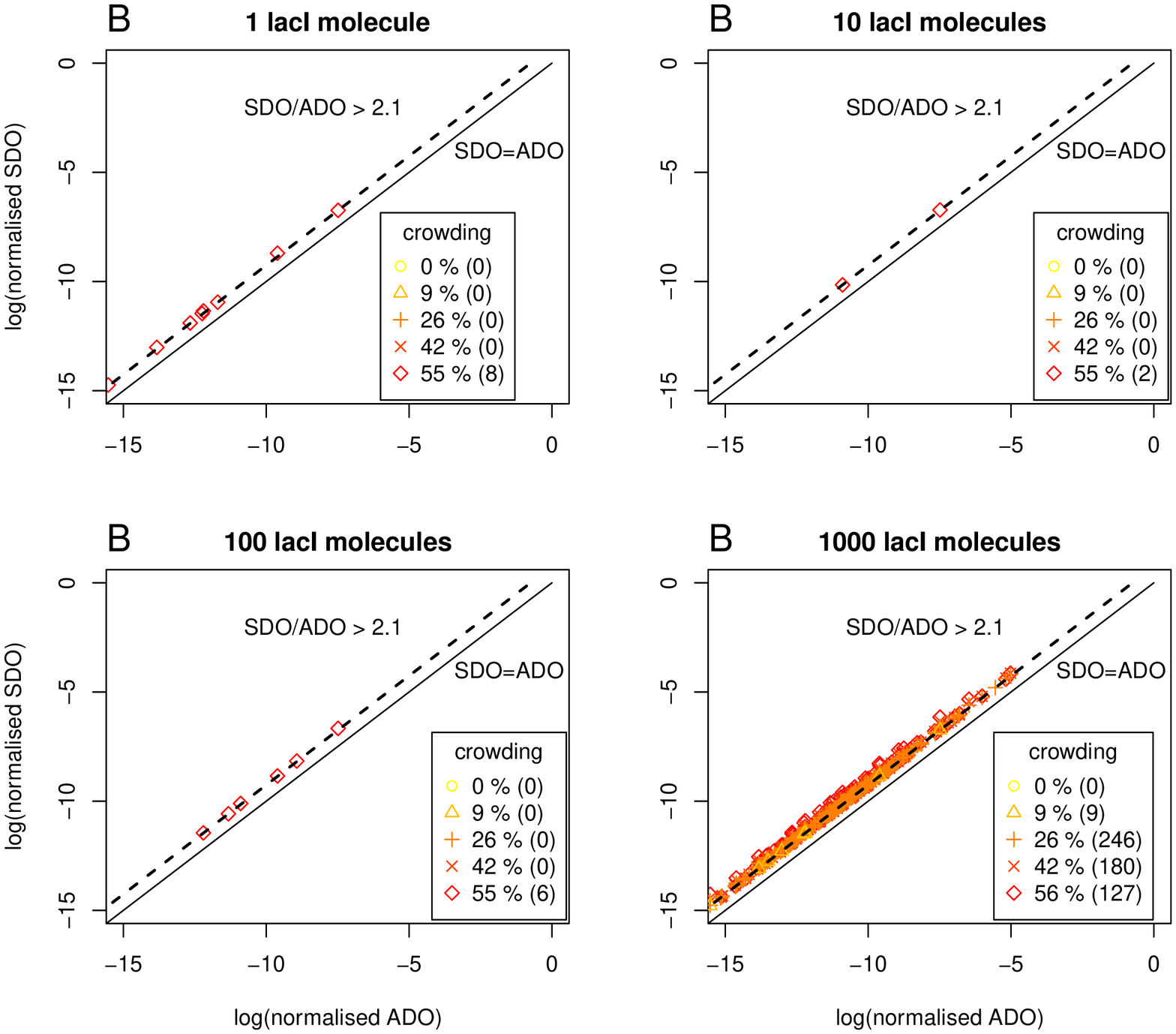}
  \end{center}
\caption[Significant deviations between ADO and SDO]{\justifying{\bf Significant deviations between ADO and SDO}. This is a the same as Figure 3 in the main manuscript, except that in this Figure we did not consider any affinity cut-off and plotted only sites where the occupancy in the simulations is at least $2.1$ times higher than that predicted by the affinity. The number in the parentheses in the legend represents the total number of sites that display an SDO at least $2.1$ times higher than the ADO for each particular case. In each panel, the abundance of lacI is kept constant and the crowding on the DNA is increased from $0\%$ to $55\%$. The level of crowding on the DNA (implemented through the abundance of non-cognate TF) influences the number of sites that display significant differences between occupancy and affinity. We considered four cases with respect to the number of lacI molecules: \onel\ $1$, \twol\ $10$, \threel\ $100$ and \fourl\ $1000$.}
\label{fig:OccupancyAffinityCrowdingScatterplotSignificant}
\end{figure}

Figure \ref{fig:OccupancyAffinityCrowdingScatterplotSignificant} shows that the sites where SDO differs significantly from ADO are medium and low affinity sites.

\newpage
\section{Generating the \emph{in silico} ChIP profile} \label{seq:appendixChIPprofile}

The R code that generates the \emph{in silico} ChIP profile (see below) is an implementation of the method described in \citep{kaplan_2011}.

\begin{quote} 
\begin{verbatim}
generateChIPProfile <- function(input.vec, mean, sd, smooth = NULL) {
    var = sd^2
    shp = mean^2/var
    scl = var/mean
    l = length(input.vec)
	
    f = dgamma(0:length(input.vec), shape = shp, scale = scl)
    F = rev(cumsum(rev(f)))
	
    peak.centres = which(input.vec > mean(input.vec))
    peaks = vector("numeric", l)
    
    for(pc in peak.centres) {
        this.peak = vector("numeric", l)
        this.peak[pc:l] = F[1:(l-pc+1)]
        this.peak[1:(pc-1)] = F[pc:2]
        peaks = peaks + this.peak * input.vec[pc]
    }
	
    if(!is.null(smooth)){
        if((smooth %% 2) == 0){smooth = smooth - 1}
        mid = round(smooth/2,0) + 1
        d = smooth - mid
        for(i in mid:(length(peaks) - d)) {
            peaks[i] = mean(peaks[max(0,(i-d)):min(length(input.vec),(i+d))])
        }
    }
	
    return(peaks)
}
\end{verbatim}
\end{quote} 

\section{Lower information content motifs} \label{seq:appendixLowerInfo}

\begin{figure}[!ht]
\begin{center}
\includegraphics[scale=0.58,angle=270]{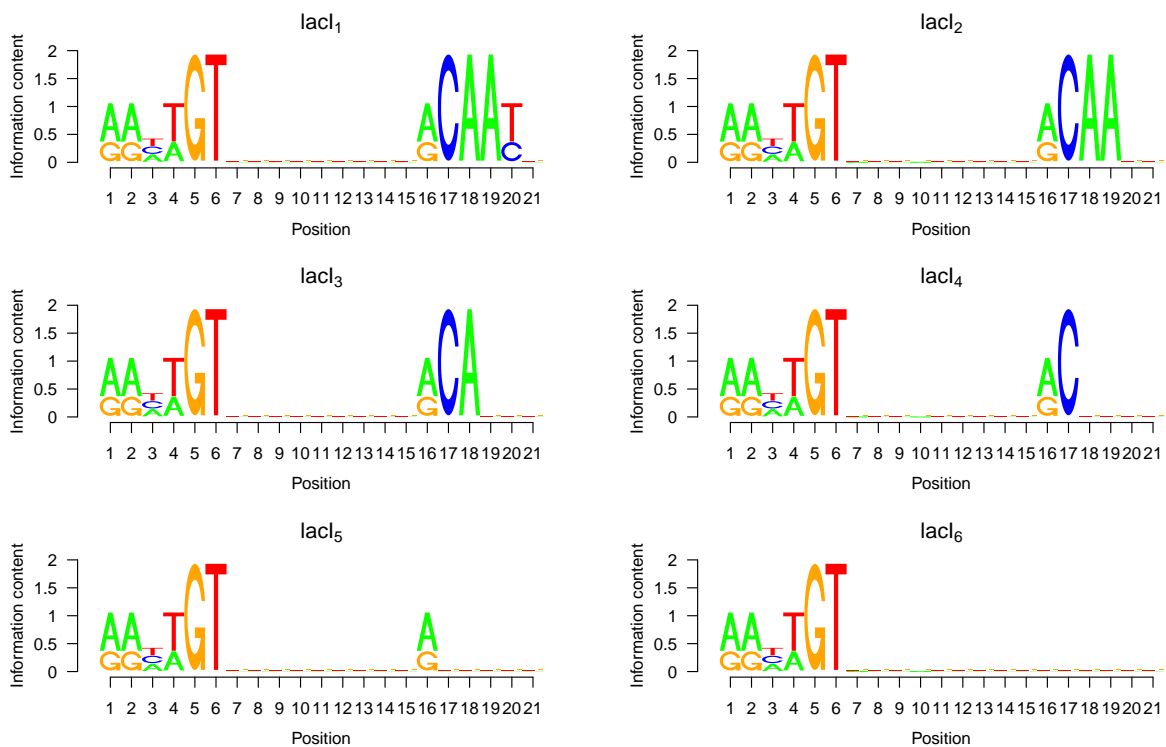}
\end{center}
\caption[Lower information content lacI motifs]{\justifying{\bf Lower information content lacI motifs}.The information content of the reduced motifs is: \one $I_{lacI_1}=15.8\ bits$, \two $I_{lacI_2}=14.7\ bits$, \three $I_{lacI_3}12.7\ bits$, \four $I_{lacI_4} = 10.7\ bits$, \five $I_{lacI_5} = 8.7\ bits$ and \six  $I_{lacI_6} = 7.7\ bits$; see Figure \ref{fig:lacI3greducedICS}.}
\label{fig:lacI3greduced}
\end{figure}

Our lacI motif has an information content of $16.9\ bits$. Hence, in order to test what is the switching limit we removed on base pair from the lacI motif and produced six new lower information content motifs; see Figure \ref{fig:lacI3greduced}.

\begin{figure}[!ht]
\begin{center}
\includegraphics[scale=0.9,angle=270]{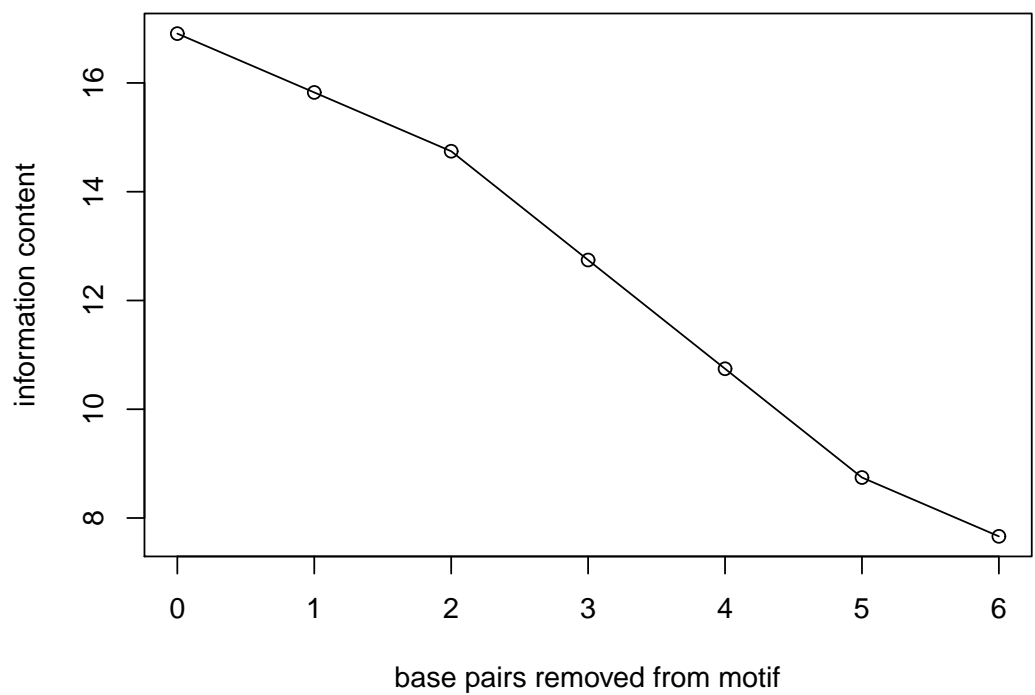}
\end{center}
\caption[Information content of the reduced lacI motifs]{\justifying{\bf Information content of the reduced lacI motifs}.}
\label{fig:lacI3greducedICS}
\end{figure}

\bibliographystyle{genomeresearch.bst}

\bibliography{grip_occupancy.bib}

\end{document}